\documentclass[12pt]{article}
\usepackage{epsf}
\usepackage{cite}
\usepackage{amsmath,amssymb}
\usepackage[dvipdf]{graphicx}
\usepackage{comment}
\usepackage{bm}
\usepackage{physics}
\usepackage{color}
\usepackage{here}
\usepackage[colorlinks,citecolor=blue]{hyperref}
\newcommand{\fR}{f_{\rm R}}

\newcommand{\MR}{M_{\mathrm R}}
\newcommand{\rdec}{r_{\rm dec}}

\makeatletter

\@addtoreset{equation}{section}
\makeatother

\begin{document}

\begin{titlepage}

\begin{center}

{\Large \bf Sneutrinos as two inflatons and curvaton \vspace{2mm} \\ 
and leptogenesis} 

\vskip .75in

{\large
Tomo~Takahashi$\,^1$, Toshifumi Yamada$\,^2$ and Shuichiro Yokoyama$\,^{3,4}$
}

\vskip 0.25in

{\em
$^{1}$Department of Physics, Saga University, Saga 840-8502, Japan  \vspace{3mm} \\
$^{2}$Department of Physics, Yokohama National University, Yokohama, 240-8501, Japan \vspace{2mm} \\
$^{3}$Kobayashi-Maskawa Institute, Nagoya University, Aichi 464-8602, Japan \vspace{3mm} \\
$^{4}$Kavli Institute for the Physics and Mathematics of the Universe (WPI), \vspace{1mm} \\ The University of Tokyo, Kashiwa 277-8583, Japan  
}

\end{center}
\vskip .5in

\begin{abstract}
We argue that sneutrinos can be embedded in a multi-field inflation framework where two inflatons and a curvaton simultaneously contribute to primordial fluctuations, which is consistent with current constraints on the spectral index and the tensor-to-scalar ratio from Planck and BICEP/Keck 2018. We also show that the same framework can also explain the baryon asymmetry of the Universe via leptogenesis realized by the decay of the lightest sneutrino. We investigate the parameter range for the scenario to work such as that of sneutrino masses. In particular, we show that the tensor-to-scalar ratio should be larger than $10^{-4}$ for a successful scenario. 
\end{abstract}

\end{titlepage}

\renewcommand{\thefootnote}{\#\arabic{footnote}}
\setcounter{footnote}{0}

\section{Introduction} \label{sec:intro}

The inflationary universe is an established paradigm describing the early stage of our Universe, but its actual mechanism remains to be elucidated. Inflation models can be tested by observing the primordial fluctuations in cosmic microwave background (CMB), large-scale structure, and so on. In particular, many models are now excluded due to observational constraints from Planck and BICEP/Keck 2018 on the spectral index of the power spectrum of the curvature perturbation, $n_s$, and the tensor-to-scalar ratio, $r$ \cite{Planck:2018jri,BICEP:2021xfz}. In particular, chaotic inflation models with any monomial power law are now excluded. 

However, we can also envisage inflationary models where multiple fields are involved in the inflationary dynamics and primordial fluctuations, and in such a framework, the predictions for the spectral index and the tensor-to-scalar ratio would be modified. Actually, some inflation models excluded as a single-field one can become viable in the multi-field framework. This issue has been studied in mixed inflaton and spectator field models \cite{Langlois:2004nn,Moroi:2005kz,Moroi:2005np,Ichikawa:2008iq,Ichikawa:2008ne,Enqvist:2013paa,Vennin:2015vfa}\footnote{
Another possible extension to alleviate inflationary models is to introduce a non-minimal coupling to gravity such as Higgs inflation \cite{Bezrukov:2007ep}.  See \cite{Kallosh:2021mnu,Cheong:2021kyc,Kodama:2021yrm} for recent work (particularly after BICEP/Keck 2018 \cite{BICEP:2021xfz}) along this direction. See also \cite{Kubota:2022pit,Hyun:2022uzc} for recent works discussing a non-minimal coupling in multi-field framework. 
}. When we focus on the quadratic chaotic inflation, in particular, the current constraints on $n_s$ and $r$ are so severe that the model cannot be rescued even if we introduce a spectator field \cite{Morishita:2022bkr}. However, with three fields, the quadratic chaotic inflation model can be consistent with current observations when two fields serve as the inflaton and one field is regarded as a spectator \cite{Morishita:2022bkr}. Indeed, three fields with all quadratic potential arise in models with sneutrinos and can be a natural candidate for the three-field chaotic inflation model. There have been many works in which one of the sneutrinos plays a role of the inflaton (Ref. \cite{Murayama:1992ua} is one of the pioneering works) or the curvaton \cite{Moroi:2002vx,Postma:2002et,McDonald:2003xq,Harigaya:2014bsa}, or some of them serve as an inflaton and a curvaton \cite{Senoguz:2012iz,Ellis:2013iea,Haba:2017fbi}. However, as mentioned above, we need three fields to make the quadratic chaotic inflation consistent with observational constraints in light of recent Planck and BICEP/Keck results, which is the case we focus on in this paper. 

As we will argue in this paper, there exists a range of parameter space where $n_s$ and $r$ can be consistent with constraints from Planck and BICEP/Keck 2018 \cite{Planck:2018jri,BICEP:2021xfz}, satisfying non-Gaussianity constraints from Planck \cite{Planck:2019kim}. Furthermore, we can also construct a model which provides the baryon asymmetry of the Universe via non-thermal leptogenesis in the context of the sneutrinos (see, e.g., \cite{Murayama:1992ua,Murayama:1993em,Covi:1996wh,Hamaguchi:2001gw}). We investigate the parameter space of sneutrinos for a successful scenario where primordial fluctuations are consistent with Planck and BICEP/Keck 2018 as well as the right amount of baryon asymmetry is realized. 

The structure of this paper is the following. In the next section, we first describe the setup of our model. Then in Section~\ref{sec:primordial_fluc}, we discuss primordial fluctuations in the sneutrino inflaton and curvaton model, particularly focusing on the predictions for $n_s$ and $r$ in accordance with Ref. \cite{Morishita:2022bkr}. We also show that there exists a parameter space where $n_s$ and $r$ can be consistent with current observational constraints from Planck and BICEP/Keck 2018.  In Section~\ref{sec:leptogenesis}, sneutrino leptogenesis in our model setup is discussed. Then in Section~\ref{sec:results}, the parameter ranges for a successful scenario are investigated. We conclude this paper in the final section.

\section{Setup} \label{sec:setup}

In this section, we briefly describe our model setup and summarize the formulas for various quantities to calculate observables such as the amplitude and the spectral index of the primordial power spectrum and the tensor-to-scalar ratio.

We consider the following potential:
\begin{equation}
\label{eq:potential}
V = \frac12 M_\phi^2 \phi^2 + \frac12 M_\chi^2 \chi^2 + \frac12 M_\sigma^2 \sigma^2 \,,
\end{equation}
where $\phi$ and $\chi$ play the role of the inflaton and $\sigma$ works as a curvaton. $M_\phi, M_\chi$ and $M_\sigma$ are the masses of $\phi, \chi$ and $\sigma$, respectively. This setup corresponds to the hierarchical-triple model discussed in \cite{Morishita:2022bkr}. In the following, we assume $M_\phi \gtrsim M_\chi \gg M_\sigma$.
We identify the three fields with the supersymmetric partners of gauge-singlet neutrinos that generate the tiny active neutrino masses via the seesaw mechanism~\cite{Minkowski:1977sc,Yanagida:1979as,Glashow:1979nm,Mohapatra:1979ia}.
The part of the action describing the masses and interactions of the gauge-singlet neutrinos and sneutrinos is
\begin{align} 
-S = \int{\rm d}^4x \sqrt{-g}
&\left[\frac{1}{2}M_\phi\, N_\phi^T i\sigma_2 N_\phi +\frac{1}{2}M_\chi\, N_\chi^T i\sigma_2 N_\chi
+\frac{1}{2}M_\sigma\, N_\sigma^T i\sigma_2 N_\sigma +{\rm H.c.}\right.
\nonumber\\
&+\frac{1}{2}M_\phi^2\phi^2 + \frac{1}{2}M_\chi^2\chi^2 + \frac{1}{2}M_\sigma^2\sigma^2 +{\rm H.c.}
\nonumber\\
&+ h_{\phi\alpha} \, H_u N_\phi^{T} i\sigma_2 L_\alpha + h_{\chi\alpha} \, H_u N_\chi^{T} i\sigma_2 L_\alpha
+ h_{\sigma\alpha} \, H_u N_\sigma^{T} i\sigma_2 L_\alpha +{\rm H.c.}
\nonumber\\
&\left.+\vert M_\phi \phi + h_{\phi\alpha}H_u \widetilde{L}_\alpha\vert^2
+\vert M_\chi \chi + h_{\chi\alpha}H_u \widetilde{L}_\alpha\vert^2
+\vert M_\sigma \sigma + h_{\sigma\alpha}H_u \widetilde{L}_\alpha\vert^2 \right].
\label{action}
\end{align}
Here $N_\phi, N_\chi, N_\sigma$ denote the gauge-singlet neutrinos that are supersymmetric partners of $\phi,\chi,\sigma$, respectively.
$L_\alpha$ denotes the Standard Model lepton doublets with $\alpha$ labeling the flavors, $\widetilde{L}_\alpha$ their supersymmetric partners, and $H_u$ the up-type Higgs field of the supersymmetric Standard Model.
$h_{i\alpha}$ $(i=\phi,\chi,\sigma)$ are the neutrino Dirac Yukawa couplings.
The action Eq.~(\ref{action}) can be derived from the K\"ahler potential and superpotential given in Appendix~A.

The active neutrino mass matrix, $M_\nu$, is generated via the seesaw mechanism as
\begin{align}
    (M_\nu)_{\alpha\beta} = \frac{v_u^2}{2} \left( h_{\phi\alpha} \frac{1}{M_\phi} h_{\phi\beta} + h_{\chi\alpha} \frac{1}{M_\chi} h_{\chi\beta} + h_{\sigma\alpha} \frac{1}{M_\sigma} h_{\sigma\beta}\right) \,,
\end{align}
where $v_u=\sqrt{2}\langle H_u\rangle$ with $\langle H_u\rangle$ the VEV of the up-type Higgs field.
The neutrino Dirac Yukawa coupling can be parametrized as~\cite{Casas:2001sr}
\begin{align}
    \begin{pmatrix}
        h_{\sigma\alpha} \\
        h_{\chi\alpha} \\
        h_{\phi\alpha}
        \end{pmatrix}
        = \frac{\sqrt{2}}{v_u}\begin{pmatrix}
        \sqrt{M_\sigma} & 0 & 0 \\
        0 & \sqrt{M_\chi} & 0 \\
        0 & 0 & \sqrt{M_\phi}
        \end{pmatrix}
        R_{3\times3}
        \begin{pmatrix}
        \sqrt{m_1} & 0 & 0 \\
        0 & \sqrt{m_2} & 0 \\
        0 & 0 & \sqrt{m_3}
        \end{pmatrix}
        U_{\nu} \,,
        \label{casas-ibarra}
\end{align}
 where $R_{3\times3}$ is an arbitrary complex orthogonal matrix, $m_1,m_2,m_3$ denote the mass eigenvalues of the active neutrinos, and $U_{\nu}$ denotes the neutrino flavor mixing matrix, which is almost unitary.
It follows that the decay rates of the inflatons and curvaton at the tree level are respectively expressed as
\begin{align}
\label{eq:decayrate}
    \Gamma_\phi &= \frac{4 M_\phi^2}{8\pi v_u^2}
    \left[R_{3\times3}\begin{pmatrix}
        m_1 & 0 & 0 \\
        0 & m_2 & 0 \\
        0 & 0 & m_3
        \end{pmatrix}R_{3\times3}^\dagger\right]_{33},
\\
   \Gamma_\chi &= \frac{4 M_\chi^2}{8\pi v_u^2}
   \left[R_{3\times3}\begin{pmatrix}
        m_1 & 0 & 0 \\
        0 & m_2 & 0 \\
        0 & 0 & m_3
        \end{pmatrix}R_{3\times3}^\dagger\right]_{22},
\\
    \Gamma_\sigma &= \frac{4 M_\sigma^2}{8\pi v_u^2}
    \left[R_{3\times3}\begin{pmatrix}
        m_1 & 0 & 0 \\
        0 & m_2 & 0 \\
        0 & 0 & m_3
        \end{pmatrix}R_{3\times3}^\dagger\right]_{11}.
\end{align}
Thus, once the matrix $R_{3\times3}$ is fixed, the decay rates at the tree level can be related to the masses of the active neutrinos. Furthermore, as we will see later, by properly choosing $R_{3\times3}$
(s)lepton asymmetry can arise from the interference of the one-loop diagrams~\cite{Murayama:1992ua,Covi:1996wh}.

\section{Primordial fluctuations and observational constraints} \label{sec:primordial_fluc}

In this section, we summarize the properties of primordial fluctuations generated in the three-field model of inflation, where the potential is given by Eq.~\eqref{eq:potential} with $\phi$ and $\chi$ being inflatons and $\sigma$ being a curvaton. For the details of this three-field model with quadratic potentials, see \cite{Morishita:2022bkr}. Note that all of the following expressions
are obtained at the leading order in the slow-roll approximation.

The power spectrum of the curvature perturbations in this model is formally given by
\begin{equation}
\label{eq:P}
{\cal P}_\zeta (k) = {\cal P}_\zeta^{\rm (inf)} + {\cal P}_\zeta^{\rm (cur)} \,,
\end{equation}
where ${\cal P}_\zeta^{\rm (inf)}$ and $ {\cal P}_\zeta^{\rm (cur)}$ are the ones generated from the inflaton and curvaton sector, respectively. Since we assume that both $\phi$ and $\chi$ play the role of the inflaton,  ${\cal P}_\zeta^{\rm (inf)}$ can be given, using the $\delta N$ formalism \cite{Starobinsky:1985ibc,Sasaki:1995aw}, as \cite{Morishita:2022bkr}
\begin{equation}
\label{eq:P_inf}
{\cal P}_\zeta^{\rm (inf)} = {\cal P}_\zeta^{(\phi)} + {\cal P}_\zeta^{(\chi)} 
 = \left( \frac{\partial N_\ast}{\partial \phi} \right)^2  \left|\delta \phi_\ast\right|^2 + \left( \frac{\partial N_\ast}{\partial \chi} \right)^2  \left|\delta \chi_\ast\right|^2  
 = \frac{N_\ast}{M_{\rm pl}^2} \left( \frac{H_\ast}{2\pi} \right)^2 \,,
\end{equation}
where $H_\ast$ and $N_\ast$ are the Hubble parameter and the number of $e$-folds at the time when the reference scale $k_\ast$ exited the horizon. The quantities with a subscript $\ast$ are those evaluated at this epoch. Since we assume that $\sigma$ is the curvaton field, its energy density is negligible during inflation, and hence $H_\ast$ can be given by
\begin{equation}
\label{eq:infHubble}
H_\ast^2 =  \frac{1}{6M_{\rm pl}^2} \left( M_\phi^2 \phi_\ast^2 + M_\chi^2 \chi_\ast^2 \right)
= \frac{M_\phi^2 }{6} \left( \frac{\phi_\ast}{M_{\rm pl}} \right)^2 ( 1+ \MR^2 \fR^2) \,,
\end{equation}
where, in the 2nd equality, we have defined 
\begin{equation}
\label{eq:def_fR_mR}
\MR \equiv \frac{M_\chi}{M_\phi}  \qquad
\fR  \equiv \frac{\chi_\ast}{\phi_\ast} \,. 
\end{equation}
The number of $e$-folds is given by 
\begin{equation}
\label{eq:Ne}
N_\ast = \frac{1}{4 M_{\rm pl}^2} \left( \phi_\ast^2 + \chi_\ast^2 \right)  
=   \frac{1}{4} \left( \frac{\phi_\ast}{M_{\rm pl}} \right)^2  (1+ \fR^2) \,,
\end{equation}
with $\phi_\ast$ and $\chi_\ast$ being the field value at the horizon exit of the reference scale $k_\ast$. In the following, we fix the number of $e$-folds at the reference scale as $N_\ast = 60$. Once we fix $\fR$ in addition to $N_\ast$, which will be done as described in the following, $\phi_\ast$ can be determined. By using the above expressions, the power spectrum of the curvature perturbations generated from the inflaton can be rewritten in terms of $N_\ast$, $\MR$, $\fR$, and $M_\phi$ as
\begin{equation}
\label{eq:P_inf_re}
 {\cal P}_\zeta^{\rm (inf)} = \frac{N_\ast^2}{6 \pi^2} 
 \left( \frac{M_\phi}{M_{\rm pl}} \right)^2 \frac{1+ \MR^2 \fR^2}{1+ \fR^2}\,.
\end{equation}

$ {\cal P}_\zeta^{\rm (cur)}$ is the curvature power spectrum generated by the curvaton, and hence it can be given as 
\begin{equation}
\label{eq:P_cur}
 {\cal P}_\zeta^{\rm (cur)} = \frac{4 \rdec^2}{9 \sigma_\ast^2}  \left( \frac{H_\ast}{2\pi} \right)^2 \,.
\end{equation}
Here $\rdec$ roughly represents the ratio of the energy density of the curvaton to the total one at the curvaton decay, which can be given by 
\begin{equation}
\label{eq:r_dec}
\rdec = 
\begin{cases}
\min \left[  1, \,\, \left( \displaystyle\frac{\sigma_\ast}{M_{\rm pl}} \right)^2 \sqrt{\displaystyle\frac{M_\sigma}{\Gamma_\sigma}} \right] & {\rm for}~M_\sigma < \min \left[\Gamma_\chi, \Gamma_\phi \right] \cr\cr
\min \left[ 1, \,\, \left( \displaystyle\frac{\sigma_\ast}{M_{\rm pl}} \right)^2 \sqrt{\displaystyle\frac{\min \left[\Gamma_\chi, \Gamma_\phi \right]}{\Gamma_\sigma}} \right] & {\rm for}~M_\sigma > \min \left[\Gamma_\chi, \Gamma_\phi \right]
\end{cases}~ \,.
\end{equation}
If $ M_\sigma < \min \left[\Gamma_\chi, \Gamma_\phi \right]$, the curvaton starts to oscillate after the decays of both inflatons, i.e., during radiation-dominated era. On the other hand, if $ M_\sigma > \min \left[\Gamma_\chi, \Gamma_\phi \right]$, the curvaton begins to oscillate while either $\phi$ or $\chi$ is still oscillating, when the universe is dominated by the energy density of oscillating $\phi$ or $\chi$, i.e., the universe is in the matter(-like)-dominated era.

By using Eqs.~\eqref{eq:P_inf_re} and \eqref{eq:P_cur}, the power spectrum for the total curvature perturbation at the reference scale $k_\ast$ is given by
\begin{equation}
\label{eq:P_tot}
{\cal P}_\zeta (k_\ast) = \frac{N_\ast^2}{6 \pi^2} 
 \left( \frac{M_\phi}{M_{\rm pl}} \right)^2 \frac{1+ \MR^2 \fR^2}{1+ \fR^2} (1+R) \,,
\end{equation}
with
\begin{equation}
\label{eq:R}
R \equiv  \frac{ {\cal P}_\zeta^{\rm (cur)}}{{\cal P}_\zeta^{\rm (inf)}} = \frac{4}{9} \frac{\rdec^2}{N_\ast} \left( \frac{\sigma_\ast}{M_{\rm pl}} \right)^{-2} \,.
\end{equation}
The spectral index and the tensor-to-scalar ratio in this model are given by 
\begin{eqnarray}
\label{eq:spi}
n_s -1  & = & - 2 \epsilon_\ast - \frac1{1+R} \frac1{N_\ast}  + \frac{R}{1+R} \frac{2M_\sigma^2}{3 H_\ast^2} \,,  \\
\label{eq:tsr}
r  & = &   \frac1{1+R} \frac8{N_\ast} \,.
\end{eqnarray}
Here $\epsilon_\ast$ is the slow-roll parameter at the horizon exit which can be expressed as~\cite{Morishita:2022bkr} 
\begin{equation}
\label{eq:epsilon}
\epsilon_\ast = \frac1{2N_\ast} \left[
1 + \frac{\fR^2 (1 - \MR^2)^2}{(1+ \MR^2 \fR^2)^2}
\right] \,.
\end{equation}
Thus, once the number of $e$-folds is fixed, $R$ can completely control the tensor-to-scalar ratio.

In the pure curvaton scenario (i.e., the curvaton is totally responsible for the curvature perturbation), the non-linearity parameter for the local type $f_{\rm NL}$ can be written as (see, e.g., \cite{Sasaki:2006kq}) 
\begin{equation}
\label{ }
\frac65 f_{\rm NL}^{\rm (curvaton)} = \frac{3}{2 \rdec} - 2 - \rdec \,.
\end{equation}
By assuming that the inflaton sector gives negligible non-Gaussianities, which we believe that this is the case in our setup, $f_{\rm NL}$ in our three-field model is given by 
\begin{equation}
\label{eq:nonG}
\frac65 f_{\rm NL} = \left( \frac{R}{1+R}  \right)^2 \frac65 f_{\rm NL}^{\rm (curvaton)}   =\left( \frac{R}{1+R}  \right)^2 \left[ \frac{3}{2 \rdec} - 2 - \rdec \right] \,. 
\end{equation}

With the setup and the predictions for the observables described above, now we discuss the conditions (constraints) for the model parameters for a successful model consistent with the recent constraints on $n_s$ and $r$ given by Planck+BICEP/Keck 2018 result \cite{BICEP:2021xfz}.
%
\begin{figure}[htbp]
\centering
  \includegraphics[width=15cm]{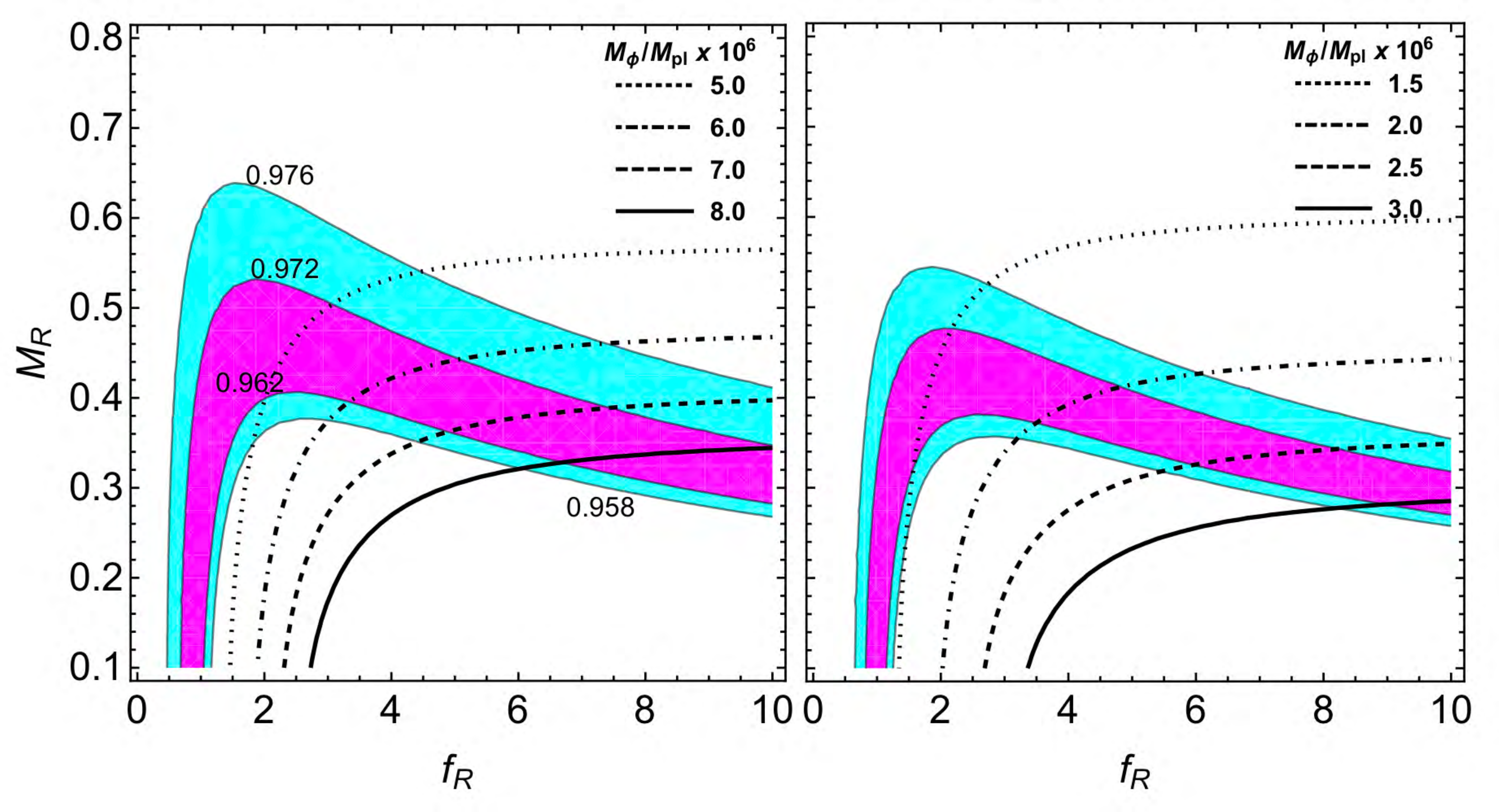}
\caption{\label{fig:fR_mR_r3em3} $1\sigma$ (magenta) and $2\sigma$ (cyan) allowed region from the measurement of $n_s$  in the $\fR$--$\MR$ plane.  Contours of $M_\phi$ are also depicted. The other parameters are fixed as $N_\ast = 60$ and $r = 3 \times 10^{-2}$ (left) and $r = 3 \times 10^{-3}$ (right).}
\end{figure}
%

\bigskip\bigskip\bigskip
\noindent 
$\bullet$  {\bf Spectral index}  \vspace{2mm} \\
In the simple curvaton scenario,
the mass of the curvaton is assumed to be negligibly small compared to that of the inflaton, and hence $M_\sigma^2/H_\ast^2$ term in the expression of the spectral index can be neglected. We employ this assumption and then the spectral index is simply given by
\begin{equation}
    n_s - 1 = - 2 \epsilon_\ast - \frac{1}{1+R} \frac{1}{N_\ast}~.
    \label{eq:simplens}
\end{equation}
Thus, once the number of $e$-folds and the tensor-to-scalar ratio (or $R$; see Eq.~\eqref{eq:tsr}) are fixed, the spectral index is given as a function of $\fR$ and $\MR$ through the expression of $\epsilon_\ast$ \eqref{eq:epsilon}. In Fig.~\ref{fig:fR_mR_r3em3}, we plot $1 \sigma$ (magenta) and $2 \sigma$ (cyan) allowed regions for the spectral index obtained from the Planck result \cite{Planck:2018jri} in the $\fR - \MR$ plane. Here, we take $N_\ast = 60$ and $r=3 \times 10^{-2}$ ($R \simeq 3.4$) in the left panel and  $r = 3 \times 10^{-3}$ ($R \simeq 43$) in the right panel.
For smaller $r$ (larger $R$) the second term of the right hand side in the expression \eqref{eq:simplens}
becomes negligible and the spectral index is independent of the tensor-to-scalar ratio. Furthermore, by taking the ratio of the masses of the two inflatons to be $ 0.3 \lesssim \MR \lesssim 0.5$, we can realize the spectral index that is consistent with the Planck result whatever value of $\fR$ is chosen. This means that for such a range of $\MR$ the prediction of the spectral index
is almost independent of the initial condition of the ratio of the inflatons' field values for the fixed number of $e$-folds. On the other hand, $\fR = {\cal O}(1)$ is somewhat natural, and hence we take $\fR=5$ in the following analysis. We also fix $\MR$ as $\MR =0.4$, which is consistent with the current observation of $n_s$ as mentioned above.

\bigskip\bigskip\bigskip
\noindent 
$\bullet$  {\bf Tensor-to-scalar ratio}  \vspace{2mm} \\
The recent BICEP/Keck result~\cite{BICEP:2021xfz} gives the upper bound
on the tensor-to-scalar ratio as $r_{0.05}<0.036$ (95 \% C.L.). From Eq.~\eqref{eq:tsr}, this constraint
implies
\begin{eqnarray}
R > 2.7\,~{\rm for}~N_\ast = 60.
\end{eqnarray}
This means that the curvature perturbations generated by the curvaton should be almost dominant.

\bigskip\bigskip\bigskip
\noindent 
$\bullet$  {\bf Normalization of the curvature power spectrum}  \vspace{2mm} \\
Since the amplitude of the curvature perturbation at the CMB scale is precisely measured, its power spectrum should be normalized as ${\cal P}_\zeta|_{\rm norm} \simeq 2 \times 10^{-9}$~\cite{Planck:2018vyg}.  From this requirement, by fixing $N_\ast$ and $R$ (or $r$), we can describe $M_\phi$ as a function of 
$\MR$ and $\fR$ as
\begin{equation}\label{eq:m_phi_norm}
\left( \frac{M_\phi}{M_{\rm pl}} \right)^2 = 
\frac{6 \pi^2 {\cal P}_\zeta|_{\rm norm}}{N_\ast}
 \frac{(1+\fR^2)}{(1+\MR^2 \fR^2)}\,\frac{r}{8}
\,,
\end{equation}
where we have used Eqs.~\eqref{eq:P_tot} and \eqref{eq:tsr}. In the allowed region derived from the measurement of $n_s$ from Planck shown in Fig.~\ref{fig:fR_mR_r3em3}, $M_\phi$ can only vary by a factor of $2$ in each panel. This means that once the tensor-to-scalar ratio is fixed
the mass of the inflaton is predicted only within a factor of a few.

\bigskip\bigskip\bigskip
\noindent 
$\bullet$  {\bf Non-Gaussianity}  \vspace{2mm} \\
Based on the above constraint on $R$ with
the expression for $f_{\rm NL}$ given by Eq.~\eqref{eq:nonG},
to satisfy the constraint on $f_{\rm NL}$ for the local type from Planck ($f_{\rm NL} = -0.9 \pm 5.1$) \cite{Planck:2019kim}, one needs to have $\rdec \simeq 1$, which requires, from Eq.~\eqref{eq:r_dec}, 
\begin{eqnarray}
\label{eq:const_decay}
 \left( \frac{\sigma_\ast}{M_{\rm pl}} \right)^2 \sqrt{\frac{M_\sigma}{\Gamma_\sigma}} > 1 
 && \quad{\rm for}~
 M_\sigma < \min \left[\Gamma_\chi, \Gamma_\phi \right] \cr\cr\cr
\left( \frac{\sigma_\ast}{M_{\rm pl}} \right)^2 \sqrt{\frac{\min \left[\Gamma_\chi, \Gamma_\phi \right]}{\Gamma_\sigma}} > 1 
&& \quad{\rm for}~
 M_\sigma > \min \left[\Gamma_\chi, \Gamma_\phi \right] \,.
\end{eqnarray}
Actually, one can rewrite $\sigma_\ast$ by using the tensor-to-scalar ratio $r$, with which one can translate the constraint on $f_{\rm NL}$ to the one for $r$. 

From Eq. \eqref{eq:R} with Eq. \eqref{eq:tsr}, we have 
\begin{equation}
\label{eq:sigma_ast_r}
    \left( \frac{\sigma_\ast}{M_{\rm pl}}  \right)^2
    =
    \frac{r}{18}  \left( 1-\frac{rN_\ast}{8} \right)^{-1} \rdec^2 \,,
\end{equation}
and hence the condition~\eqref{eq:const_decay} can be rewritten as \begin{eqnarray}
\label{eq:const_decay_2}
 \frac{r}{18}\left(1 - \frac{r N_\ast}{8} \right)^{-1} \sqrt{\frac{M_\sigma}{\Gamma_\sigma}} > 1 
 && \quad{\rm for}~
 M_\sigma < \min \left[\Gamma_\chi, \Gamma_\phi \right] \cr\cr\cr
 \frac{r}{18}\left(1 - \frac{r N_\ast}{8} \right)^{-1} \sqrt{\frac{\min \left[\Gamma_\chi, \Gamma_\phi \right]}{\Gamma_\sigma}} > 1 
 && \quad{\rm for}~
 M_\sigma > \min \left[\Gamma_\chi, \Gamma_\phi \right] \,,
\end{eqnarray}
where we have assumed $\rdec = 1$. This provides the constraints on model parameters and is used in Section~\ref{sec:results} when we investigate the parameter space for a successful model.

\section{Leptogenesis via sneutrino decay} \label{sec:leptogenesis}

In the previous section, we have discussed the predictions for primordial fluctuations in the three-field model where two fields act as inflaton, and the other one plays the role of the curvaton. As we mentioned in Section~\ref{sec:setup}, we identify these three fields with sneutrinos, the supersymmetric partners of gauge-singlet neutrinos. Actually, in this framework, the baryon asymmetry of the Universe can also be explained by leptogenesis~\cite{Fukugita:1986hr}. In particular, we consider the so-called non-thermal leptogenesis scenario (see, e.g.,  \cite{Murayama:1992ua,Murayama:1993em,Covi:1996wh,Hamaguchi:2001gw}), and here we assume that it is realized by the decay of the lightest sneutrino acting as the curvaton.
Interestingly, as we will show in the following, in this sneutrino framework, there exists a parameter space where a successful inflationary model consistent with the current observations and the observed baryon asymmetry of the Universe are simultaneously realized. First, we summarize the leptogenesis via the sneutrino decay in this section. Then in Section~\ref{sec:results}, we investigate the parameter space of the above-mentioned successful model. 

In our model, the lepton number asymmetry of the Universe is generated non-thermally through CP-violating decays of the curvaton-sneutrino $\sigma$. The lepton number asymmetry is then converted to that of the baryon number through the sphaleron process \cite{Klinkhamer:1984di,Kuzmin:1985mm}.
The baryon number yield, $n_B/s$ with $s$ being the entropy of the Universe, generated from the lepton number one, $n_L/s$, via the sphaleron process is given by \cite{Khlebnikov:1988sr,Harvey:1990qw,Covi:1996wh}
\begin{align}
    \frac{n_B}{s}=-\frac{8}{23}\frac{n_L}{s}.
\end{align}
The lepton number yield is related to the number density of $\sigma$ at the time of its decay as
\begin{align}
   \left. \frac{n_L}{s} = \epsilon_1 \frac{n_\sigma}{s}\right|_{t=\sigma\,{\rm decay}}.
\end{align}
Here $\epsilon_1$ is the CP-violation parameter given by
\begin{align}
    \epsilon_1 = 
    \frac{ \sum_\alpha\Gamma(\sigma \to \tilde{L}_\alpha H_u) - \sum_\alpha\Gamma(\sigma \to \tilde{L}_\alpha^\dagger H_u^\dagger) }
    { \sum_\alpha\Gamma(\sigma \to \tilde{L}_\alpha H_u) + \sum_\alpha\Gamma(\sigma \to \tilde{L}_\alpha^\dagger H_u^\dagger) } \,,
\end{align}
which can be calculated as \cite{Covi:1996wh}
\begin{align}
    \epsilon_1 = -\frac{1}{8\pi} \sum_{i=\phi,\chi} 
    \left\{\sqrt{x_i}\log\left(1+\frac{1}{x_i}\right)+\frac{2\sqrt{x_i}}{x_i-1}\right\}
    \frac{{\rm Im}[(hh^\dagger)_{\sigma i}^2]}{(hh^\dagger)_{\sigma\sigma}} \,,
    \label{eq:CPpara}
\end{align} 
 where $x_i=M_i^2/M_\sigma^2 \gg 1$ for $M_\phi \gtrsim M_\chi \gg M_\sigma$.
The number density of $\sigma$ at the time of its decay can be written as
\begin{align}
    \left.\frac{n_\sigma}{s}\right|_{t=\sigma\,{\rm decay}} \ &= \ \frac{3}{4}\frac{g_*}{g_{*S}}\frac{T_{{\rm reh}}^{(\sigma)}}{M_\sigma}
    \cr\cr
    &\ = \ \frac{3}{4}\frac{g_*}{g_{*S}}\left(\frac{90}{\pi^2g_*}\right)^{1/4}  \frac{\left(M_{\rm pl} \Gamma_\sigma\right)^{1/2}}{M_\sigma}.
\end{align}
 
We introduce the following ansatz for the complex orthogonal matrix $R_{3\times3}$ in Eq.~(\ref{casas-ibarra}). In the case when the active neutrinos have the normal mass hierarchy, we assume 
\begin{align}
\label{eq:normalansatz}
    R_{3\times3}=\begin{pmatrix}
    \cos\omega & \sin\omega & 0 \\
    -\sin\omega & \cos\omega & 0 \\
    0 & 0 & 1
    \end{pmatrix} \ \ \ \ \ \ \ \ {\rm (Normal \ hierarchy \ case)} \,,
\end{align}
where $\omega$ takes a \textit{complex} value and is considered as a free parameter. The phase of $\omega$ gives rise to non-zero $\epsilon_1$ and is responsible for the lepton number asymmetry.
The above ansatz has the feature that when $\omega \to 0$, the hierarchy of $M_\phi, M_\chi, M_\sigma$ is aligned with that of the active neutrino masses $m_1,m_2,m_3$~\cite{Haba:2017fbi}.
With $R_{3 \times 3}$ given above, the decay rates of the sneutrinos at the tree level and the CP-violation parameter $\epsilon_1$, given by Eq. \eqref{eq:CPpara} in the normal hierarchy case, are calculated as
\begin{eqnarray}
\label{eq:decay_sigma}
&&\Gamma_\sigma = \frac{4 M_\sigma^2}{8\pi v_u^2}
\frac{\left[\left(m_1 - m_2 \right) \cos 2\Re\omega + \left( m_1 + m_2 \right) \cosh 2\Im\omega \right]}{2}, \\
&&\Gamma_\chi = \frac{4 M_\chi^2}{8\pi v_u^2}
\frac{\left[\left(m_2 - m_1 \right) \cos 2\Re\omega + \left( m_1 + m_2 \right) \cosh 2\Im\omega \right]}{2}, \\
&&\Gamma_\phi = \frac{4 M_\phi^2}{8\pi v_u^2}\,m_3,\\
\label{eq:epsilon_1}
&&\epsilon_1 = -\frac{3 M_\sigma}{4\pi v_u^2}\frac{(m_2^2-m_1^2) \sin 2\Re\omega \sinh 2\Im\omega}{\left[\left(m_1 - m_2 \right) \cos 2\Re\omega + \left( m_1 + m_2 \right) \cosh 2\Im\omega \right]}.
\end{eqnarray}
The active neutrino masses, $m_2$ and $m_3$, can be written as $m_2 = \sqrt{\Delta m_{21}^2 + m_1^2}$ and $m_3 = \sqrt{\Delta m_{32}^2 + \Delta m_{21}^2 + m_1^2}$, and in the normal hierarchy case $(m_1 < m_2 < m_3)$, we have $\Delta m_{32}^2>0$. The mass differences have been measured by neutrino oscillation experiments as $\Delta m_{21}^2 = 7.53 \times 10^{-5}$ ${\rm eV}^2$ and $\left|\Delta m_{32}^2\right| = 2.44 \times 10^{-3}$ ${\rm eV}^2$ \cite{ParticleDataGroup:2020ssz}.  We use these values in the following analysis.
Thus, in this setup the lepton number asymmetry can be characterized by $m_1$, $\Re \omega$, $\Im \omega$, and $M_\sigma$.

Although we mainly discuss the normal hierarchy case in the following analysis, we can also investigate the case with the inverted mass hierarchy. In this case, we assume 
\begin{align}
    R_{3\times3}=\begin{pmatrix}
    \sin\omega & 0 & \cos\omega \\
    \cos\omega & 0 & -\sin\omega \\
    0 & 1 & 0 \\
    \end{pmatrix} \ \ \ \ \ \ \ \ {\rm (Inverted \ hierarchy \ case)} \,.
\end{align}
In the inverted hierarchy case $(m_3 < m_1 < m_2)$, we have $\Delta m_{32}^2 < 0$. The decay rate and the CP-violation parameter are calculated as 
\begin{eqnarray}
&&\Gamma_\sigma = \frac{4 M_\sigma^2}{8\pi v_u^2}
\frac{\left[\left(m_3 - m_1 \right) \cos 2\Re\omega + \left( m_3 + m_1 \right) \cosh 2\Im\omega \right]}{2} \,, \\
&&\Gamma_\chi = \frac{4 M_\chi^2}{8\pi v_u^2}
\frac{\left[\left(m_1 - m_3 \right) \cos 2\Re\omega + \left( m_1 + m_3 \right) \cosh 2\Im\omega \right]}{2} \,, \\
&&\Gamma_\phi = \frac{4 M_\phi^2}{8\pi v_u^2}\,m_2 \,, \\
&&\epsilon_1 = -\frac{3 M_\sigma}{4\pi v_u^2}\frac{(m_1^2-m_3^2) \sin 2\Re\omega \sinh 2\Im\omega}{\left[\left(m_3 - m_1 \right) \cos 2\Re\omega + \left( m_3 + m_1 \right) \cosh 2\Im\omega \right]} \,.
\end{eqnarray}
The results of the analysis in the inverted hierarchy case are shown in Appendix~C. It should be mentioned, though, that the results for the normal and inverted mass hierarchy cases are qualitatively the same.

\section{Sneutrinos as multi-field inflation and leptogenesis \label{sec:results}}

Now in this section, we discuss which parameter space can successfully realize leptogenesis while the predictions for the spectral index $n_s$ and the tensor-to-scalar ratio $r$ are consistent with Planck and BICEP/Keck 2018 results. 

Indeed there are many parameters in our setup: $M_\phi,\,M_\chi ({\rm or}~\MR = M_\chi/M_\phi),\,M_\sigma$ (the sneutrino masses), $\phi_\ast,\,\chi_\ast ({\rm or}~\fR = \chi_\ast/\phi_\ast),\,\sigma_\ast$ (the initial amplitudes of the sneutrinos), $m_{\rm lightest}$ (the lightest active neutrino mass, which corresponds to $m_1$ in the normal mass hierarchy case), $\Re \omega, \Im \omega$ (the complex angle in the $3\times 3$ rotation matrix $R_{3\times 3}$). However, as we have discussed in Sec.~\ref{sec:primordial_fluc}, several of them can be fixed by the constraints on primordial curvature perturbation and some reasonable assumptions. 

To be consistent with the constraint on $n_s$ for any initial conditions in the inflaton field space (corresponding to the choice of $\fR$), the mass ratio $\MR (= M_\chi / M_\phi)$ should be taken to be $0.3 \lesssim \MR  \lesssim 0.5$, and we fix it as $\MR = 0.4$ in the following analysis. In addition, we take the amplitude ratio  $\fR ( = \chi_\ast / \phi_\ast)$ as $\fR = 5$ since $n_s$ is not affected much by the value of $\fR$ for $\fR > {\cal O}(1)$ as shown in Fig.~\ref{fig:fR_mR_r3em3} and having almost similar amplitudes for $\phi$ and $\chi$ is reasonable. Since we assume $\sigma$ to be a curvaton, the number of $e$-folds is determined by $\phi_\ast$ and $\chi_\ast$ (see Eq.~\eqref{eq:Ne}) and hence by fixing $N_\ast$\footnote{
In principle, in our setup the decay rates of the three fields are also given by the sneutrino masses, $m_{\rm lightest}$, $\Re \omega$ and $\Im \omega$, and hence the required number of $e$-folds can be expressed as a function of these parameters. However, the change of the number of $e$-folds is only a few percent at most for the parameter ranges we consider and is not expected to significantly affect the results.
} and $\fR$, the values of $\phi_\ast$ and $\chi_\ast$ are determined. 
Although $\sigma_\ast$ can be regarded as a free parameter, once $N_\ast$ is given, $\sigma_\ast$ can be determined by the tensor-to-scalar ratio $r$ by  requiring $\rdec = 1$ from the constraint on the non-Gaussianity (see Eq. \eqref{eq:sigma_ast_r}). Therefore we regard $r$ as a model parameter instead of $\sigma_\ast$. 
Similarly, once $N_\ast$ is given, $M_\phi$ is mostly determined given the tensor-to-scalar ratio $r$ although it depends weakly on $\fR$ and $\MR$ (see Eq. \eqref{eq:m_phi_norm}).
Regrading the mass of the lightest neutrino $m_{\rm lightest}$, it should be at least smaller than the cosmological bound for the sum of neutrino masses $\sum m_\nu < 0.12~{\rm eV}$ \cite{Planck:2018vyg}\footnote{
In the light of the so-called Hubble constant tension, the cosmological bound on neutrino masses may be affected. In the framework that the Hubble constant tension can be resolved, the constraint on neutrino masses gets less severe \cite{Sekiguchi:2020igz}.
}.
As we discuss below, the value of $\Re \omega$ is chosen such that the CP violation parameter $\epsilon_1$ is maximized for a given $\Im \omega$.

\begin{figure}[H]
\centering
  \includegraphics[width=16cm]{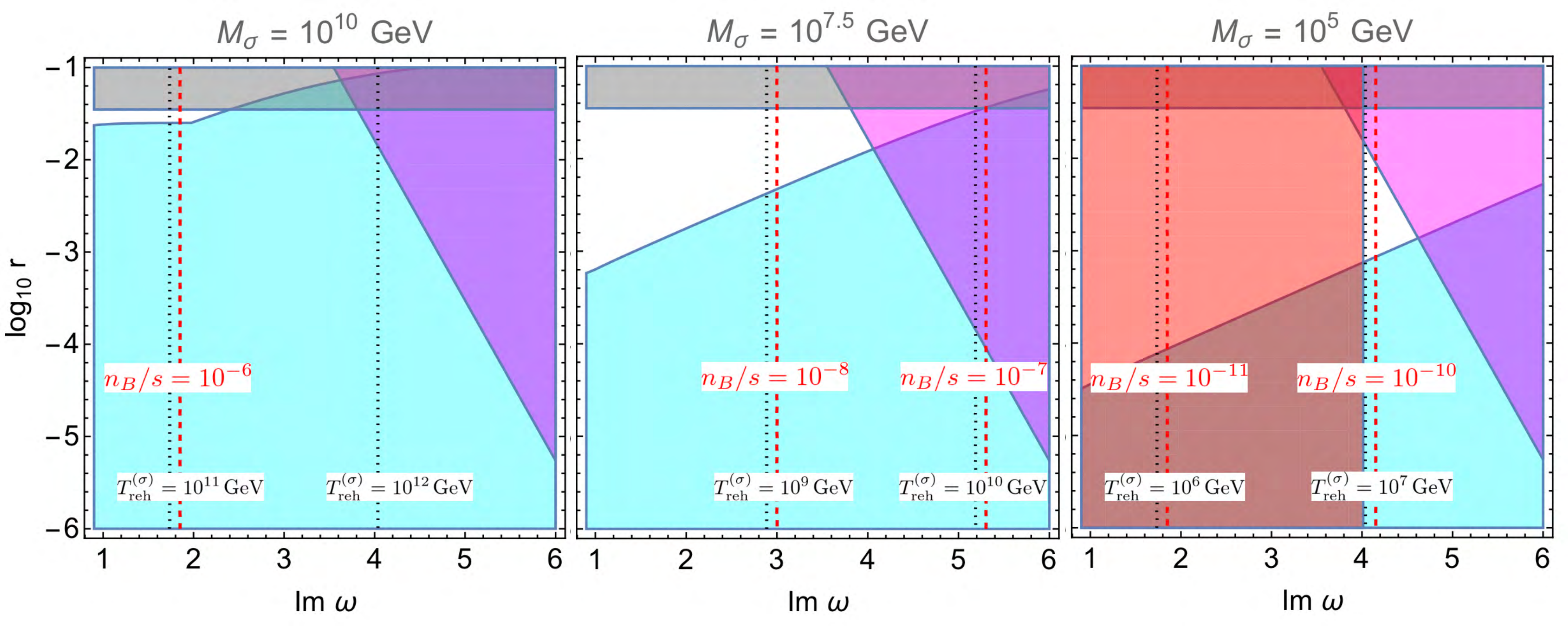}
\caption{\label{fig:const_Imw_r} Constraints on the $\Im \omega$ - $r$ plane. From the left to the right, we change the mass of the curvaton (the lightest sneutrino) as $M_\sigma = 10^{10}$, $10^{7.5}$, and $10^5$~${\rm GeV}$. The grey region is excluded by the BICEP/Keck 2018 result on $r$. In the magenta region, the perturbativity of the Yukawa coupling is broken down. The cyan region is excluded by the requirement that $\rdec = 1$ coming from the constraint on the primordial non-Gaussianity. The vertical red dashed lines are corresponding to the contours of the baryon-to-entropy ratio, $n_B / s$, and the black-dotted lines are that of the reheating temperature by the curvaton, $T_{\rm reh}^{(\sigma)}$. The red region is that inaccessible to account for the present value of the baryon-to-entropy ratio. Since we take the value of $\Re \omega$ such that $\epsilon_1$ is maximized, we can always realize the observed baryon asymmetry when $n_B/s > 8.7 \times 10^{-11}$ by tuning $\Re \omega$.}
\end{figure}

\begin{figure}[H]
\centering
  \includegraphics[width=16cm]{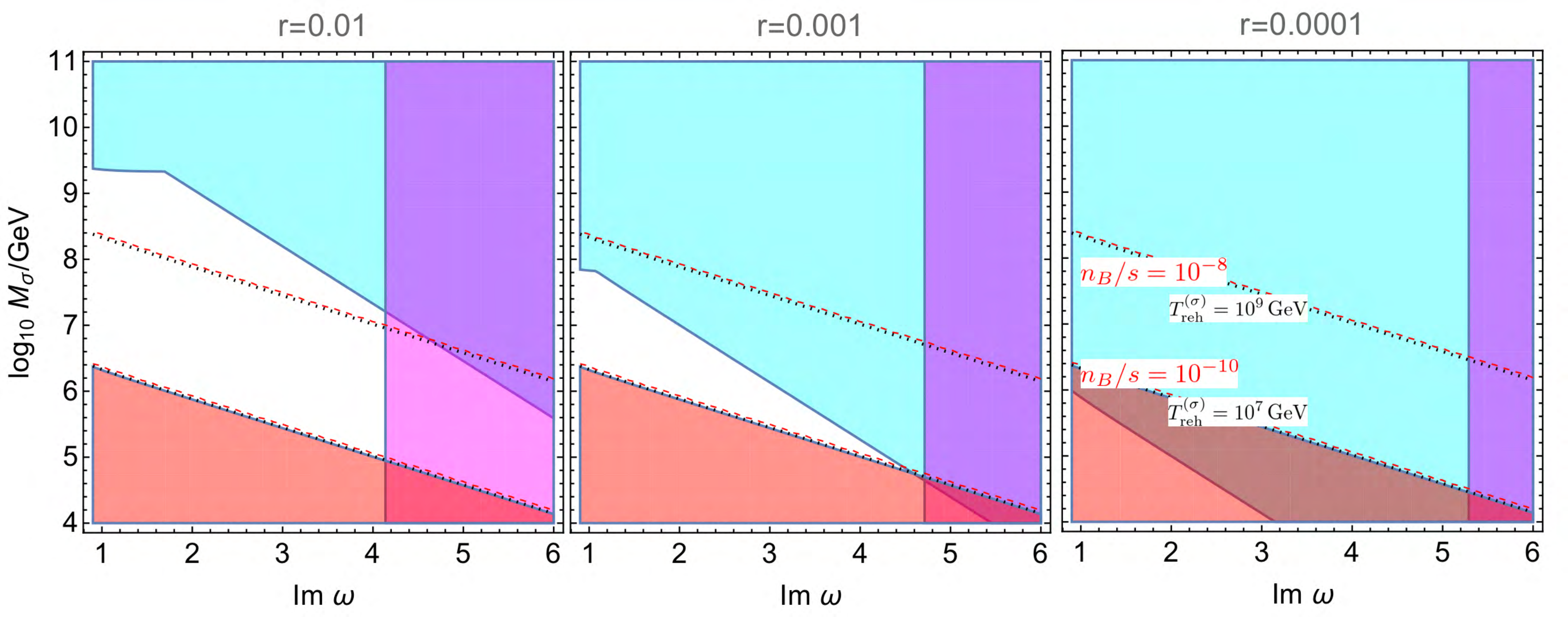}
\caption{\label{fig:const_Imw_Msigma} Constraints on the $\Im \omega$ - $M_\sigma$ plane. From the left to the right, we change the tensor-to-scalar ratio as $r=0.01$, $0.001$, and $0.0001$. The magenta region indicates the breakdown of the perturbativity of the Yukawa coupling.
The cyan region is excluded by the requirement that
$\rdec = 1$ coming from the constraint on the primordial non-Gaussianity. In each panel, the red dashed lines correspond to $n_B/s = 10^{-8},$ and $10^{-10}$, from the top to the bottom, and
the black dotted lines correspond to $T_{\rm reh}^{(\sigma)}=10^9$, and $10^{7}$ GeV, from the top to the bottom. Since we take the value of $\Re \omega$ such that $\epsilon_1$ is maximized, we can always realize the observed baryon asymmetry when $n_B/s > 8.7 \times 10^{-11}$ by tuning $\Re \omega$. The red region is that inaccessible to account for the present value of the baryon-to-entropy ratio.}
\end{figure}

Then we are left with 3 main free parameters: $M_\sigma, r $ and $\Im \omega$, and a sub-parameter: $m_{\rm lightest}$. These parameters are restricted by requiring that (i) we have the right amount of baryon asymmetry, (ii) the perturbative approach should not be broken, (iii) the tensor-to-scalar ratio should be consistent with the observational bound from Planck+BICEP/Keck 2018, and (iv) the constraint on (local-type) non-Gaussianity should be satisfied. In Figs.~\ref{fig:const_Imw_r} and \ref{fig:const_Imw_Msigma}, excluded regions from  (i)--(iv) are shown in the $\Im \omega$--$r$ plane with $M_\sigma$ being fixed, and in the $\Im \omega$--$M_\sigma$ with $r$ being fixed, respectively. In the figures, we fix $m_{\rm lightest} = 10^{-5}\,{\rm eV}$. The colored regions show excluded ones, and hence the white region is allowed. 
%
%
%
In the following, first we give some detailed descriptions for the requirements (i)--(iv), and then the discussion on the implications of the results shown in  Figs.~\ref{fig:const_Imw_r} and \ref{fig:const_Imw_Msigma} follows.

\bigskip\bigskip\bigskip
\noindent 
{\bf (i) Baryon asymmetry} \vspace{2mm} \\
$\omega$ in the matrix $R_{3\times 3}$ is regarded as a free parameter in the model, which affects the size of $\epsilon_1$. To investigate how the value of $\epsilon_1$ depends on $\omega$, we introduce a  function ${\mathcal F}_{\epsilon_1}$ as
\begin{eqnarray}
&&{\mathcal F}_{\epsilon_1} (m_1,m_2,\Re \omega, \Im \omega):=
\frac{(m_1 + m_2) \sin 2\Re\omega \sinh 2\Im\omega}{\left[\left(m_1 - m_2 \right) \cos 2\Re\omega + \left( m_1 + m_2 \right) \cosh 2\Im\omega \right]} \,,
\label{eq:mathF_epsilon}
\end{eqnarray}
with which $\epsilon_1$ can be written as
\begin{eqnarray}
&&\epsilon_1 =  -\frac{3 M_\sigma}{4\pi v_u^2}
(m_2 - m_1)  {\mathcal F}_{\epsilon_1} (m_1,m_2,\Re \omega, \Im \omega) \,,
\end{eqnarray}
and then, the baryon-to-entropy ratio is given by 
\begin{eqnarray}
\frac{n_B}{s} \approx 1.0 \times 10^{-10}  \left(\frac{m_2 - m_1}{10^{-2}~{\rm eV}} \right)
\left(\frac{T_{\rm reh}^{(\sigma)}}{10^7~{\rm GeV}} \right) {\mathcal F}_{\epsilon_1} (m_1,m_2,\Re \omega, \Im \omega) \,,
\label{eq:baryon_reheat}
\end{eqnarray}
where we have assumed $M_\sigma \ll M_\chi$ and used $v_u = 246~{\rm GeV}$. 
%
\begin{figure}[htbp]
\centering
  \includegraphics[width=16cm]{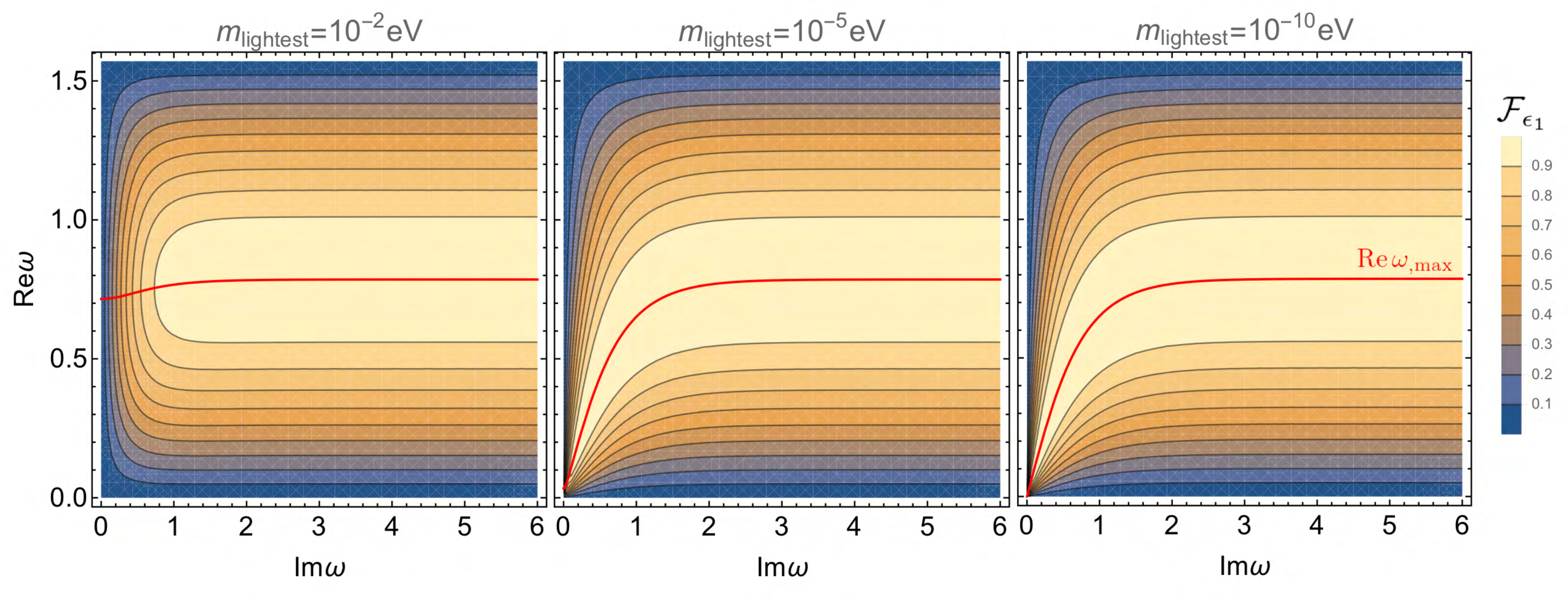}
\caption{\label{fig:contour_Fepsilon} Density contour plot of ${\mathcal F}_{\epsilon_1}$ in the $\Im \omega$ -- $\Re \omega$ plane. The mass of the lightest neutrino are taken as $m_1 = m_{\rm lightest}=10^{-2}$ (left), $10^{-5}$ (middle), and $10^{-10}$~eV (right). The red line in each panel corresponds to the analytic expression which gives the maximum value of ${\mathcal F}_{\epsilon_1}$.}
\end{figure}
%

In Fig.~\ref{fig:contour_Fepsilon}, a density contour plot of ${\cal F}_{\epsilon_1}$ is shown in the $\Im \omega$ -- $\Re \omega$ plane. As seen from the figure, when $\Im \omega \gtrsim {\cal O}(1)$, ${\cal F}_{\epsilon_1}$ scarcely depends on $\Im \omega$ and $m_{\rm lightest}$, and is almost controlled by $\Re \omega$. On the other hand, in the small $\Im \omega $ region, the value of ${\cal F}_{\epsilon_1}$ is sensitive to $\Re \omega$, $\Im \omega$, and $m_{\rm lightest}$. In either case, for a fixed $\Im \omega$, we can maximize the value of ${\cal F}_{\epsilon_1}$ by taking $\Re \omega$ as 
\begin{eqnarray}
\Re \omega_{,{\rm max}} (\Im \omega) = \frac{1}{2}\arccos{\left[ \frac{(m_2 - m_1)}{(m_2 + m_1)} \sech 2\Im \omega \right]} \,,
\label{eq:analytic_max}
\end{eqnarray}
from which we can also see that the value of $\Im \omega$ is irrelevant to the maximum ${\cal F}_{\epsilon_1}$ when $\Im \omega \gtrsim 1$. 
In the following analysis, the value of $\Re \omega$ is taken to the one given in Eq.~\eqref{eq:analytic_max} so that ${\cal F}_{\epsilon_1}$ is maximized. With this value of ${\cal F}_{\epsilon_1}$, when we obtain the baryon-to-entropy ratio $n_B/s$ larger than the observed one,  $n_B/s \simeq 8.7 \times 10^{-11}$, we can have the right amount of baryon asymmetry by tuning the value of $\Re \omega$ for a given $\Im \omega$. 

In Figs.~\ref{fig:const_Imw_r} and \ref{fig:const_Imw_Msigma}, contours of $n_B/s$ are shown with red dashed lines. Since $r$ is irrelevant to predict $n_B/s$, contours run vertically in Fig.~\ref{fig:const_Imw_r} and a larger $\Im\omega$ gives a larger $n_B/s$. Since we choose $\Re\omega$ such that ${\cal F}_{\epsilon_1}$ is maximized, for example, the right region from the line of $n_B/s = 10^{-9}$, we can always obtain the observed baryon asymmetry by tuning $\Re \omega$. On the other hand, the left region from the line of $n_B/s = 10^{-9}$, the value of $n_B/s$ is smaller than $10^{-9}$. Therefore the left region from the line of $n_B/s = 8.7 \times 10^{-11}$ is excluded since one cannot realize the observed baryon asymmetry of the Universe even for $\Re \omega$ which maximizes ${\cal F}_{\epsilon_1}$. In Fig.~\ref{fig:const_Imw_Msigma}, contours of $n_B /s = 10^{-8} $ and $10^{-10}$ are shown. As mentioned above, $r$ is irrelevant to $n_B/s$, all panels in Fig.~\ref{fig:const_Imw_Msigma} have the same contours of $n_B/s$. 
From Eq.~\eqref{eq:baryon_reheat}, the baryon-to-entropy ratio is proportional to $T_{\rm reh}^{(\sigma)}$,
which 
can be seen from the overlapping lines of $n_B/s$ and $T_{\rm reh}^{(\sigma)}$ in the figure.
$T_{\rm reh}^{(\sigma)}$ is proportional to $\Gamma_\sigma^{1/2}$
and, from Eq.~\eqref{eq:decay_sigma}, $T_{\rm reh}^{(\sigma)} \propto M_\sigma (\cosh 2 \Im \omega)^{1/2}$ for $\Im \omega \gtrsim O(1)$. 
Since we fix the values of some parameters as described at the beginning of this section, the excluded region of $n_B/s < 8.7 \times 10^{-11}$ can be roughly estimated as
\begin{eqnarray}
  \left( \frac{M_\sigma}{10^{6}~{\rm GeV}} \right)^{2} \cosh 2 \Im \omega \lesssim 12 \,.
\end{eqnarray}
One can see this dependence in Fig.~\ref{fig:const_Imw_Msigma}.
The region below the line of $n_B/s=8.7 \times 10^{-11}$ is excluded since we cannot have the right amount of baryon asymmetry even if we tune $\Re\omega$ anymore in such region.

\bigskip\bigskip\bigskip
\noindent 
{\bf (ii) Perturbativity of the Yukawa coupling} \vspace{2mm} \\
From the ansatz  
for $R_{3\times 3}$ in
the normal mass hierarchy case, given in Eq.~\eqref{eq:normalansatz}, the Yukawa coupling for $\chi$-field is given by
\begin{eqnarray}
\label{eq:hh_chichi}
(hh^\dagger)_{\chi\chi} &=& \frac{M_\chi}{v_u^2}
\left[\left(m_2 - m_1 \right) \cos 2\Re\omega  + \left( m_1 + m_2 \right) \cosh 2\Im\omega
\right]  \,.
\end{eqnarray}
Notice that $M_\chi$ is related to $r$ as
\begin{eqnarray}
M_\chi
= \MR M_{\rm pl} \left[ 
\frac{6 \pi^2 {\cal P}_\zeta|_{\rm norm}}{N_\ast}
 \frac{(1+\fR^2)}{(1+\MR^2 \fR^2)}\,\frac{r}{8} \right]^{1/2}
\,,
\end{eqnarray}
where we have used $\MR = M_\chi/M_\phi$ and 
Eq.~\eqref{eq:m_phi_norm}.  As seen from Eq.~\eqref{eq:hh_chichi}, when $\Im \omega$ becomes large, $(hh^\dagger)_{\chi\chi}$ also does. In order for the perturbative approach not to break down, we require that $(hh^\dagger)_{\chi\chi} < 1$. Regions excluded by this requirement is shown in magenta in Figs.~\ref{fig:const_Imw_r} and \ref{fig:const_Imw_Msigma}.
The excluded region of $(hh^\dagger)_{\chi\chi} > 1$ can be roughly written as 
\begin{equation}
    \left( \frac{r}{10^{-2}} \right)^{1/2} \cosh 2 \Im \omega \gtrsim 2000 \,.
\end{equation}
Since $M_\sigma$ is not relevant to this requirement, the above formula explains how the excluded region appears in Figs.~\ref{fig:const_Imw_r} and \ref{fig:const_Imw_Msigma}.

\bigskip\bigskip\bigskip
\noindent 
{\bf (iii) Tensor-to-scalar ratio}  \vspace{2mm} \\
The current upper bound on the tensor-to-scalar ratio is obtained by the BICEP/Keck 2018 data
as $r_{0.05} < 0.036$ ($95~\%$ C.L.) \cite{BICEP:2021xfz}.
The excluded region from this requirement corresponds to the upper grey band in Fig.~\ref{fig:const_Imw_r}. When we depict Fig.~\ref{fig:const_Imw_Msigma}, we assume the tensor-to-scalar ratio satisfying the constraint, and hence this requirement is irrelevant to Fig.~\ref{fig:const_Imw_Msigma}.

\bigskip\bigskip\bigskip
\noindent 
{\bf (iv) Primordial non-Gaussianity} \vspace{2mm} \\
As discussed in Sec.~\ref{sec:primordial_fluc}, the constraint on the primordial non-Gaussianity requires that Eq.~\eqref{eq:const_decay_2}
should hold. In Figs.~\ref{fig:const_Imw_r} and  \ref{fig:const_Imw_Msigma}, the cyan region corresponds to the one where this requirement is not satisfied. The behavior of the boundary of the region changes at relatively large $r$ (for example, for the case of $M_\sigma=10^{10}~{\rm GeV}$ in Fig.~\ref{fig:const_Imw_r}, the behavior changes at around upper left region). This corresponds to the transition from $M_\sigma < \min \left[ \Gamma_\chi, \Gamma_\phi \right]$ to  $M_\sigma > \min \left[ \Gamma_\chi, \Gamma_\phi \right]$ (i.e., from the 1st equation to the 2nd one in Eq.~\eqref{eq:const_decay_2}). In the parameter range we consider here, $\min \left[ \Gamma_\chi, \Gamma_\phi \right] = \Gamma_\chi$, and hence when $M_\sigma > \min \left[ \Gamma_\chi, \Gamma_\phi \right]$ holds, $\Im \omega$ dependence in $\sqrt{\min \left[ \Gamma_\chi, \Gamma_\phi \right]/{\Gamma_\sigma}}$ almost cancels out. This is the reason why $\Im \omega$ dependence disappears in the region of $M_\sigma > \min \left[ \Gamma_\chi, \Gamma_\phi \right]$, as seen from the figures.
The rough estimate for the excluded region is
\begin{eqnarray}
    \left( \frac{r}{10^{-3}} \right)^{2} \left( \frac{M_\sigma}{10^6~{\rm GeV}} \right)^{-1} \left( \cosh 2 \Im \omega \right)^{-1} \lesssim 0.004 && \quad{\rm for}~
 M_\sigma < \min \left[\Gamma_\chi, \Gamma_\phi \right] \cr\cr\cr
  \left( \frac{r}{10^{-2}} \right)^{3/2} \left( \frac{M_\sigma}{10^{10}~{\rm GeV}} \right)^{-1}  \lesssim 5 &&\quad{\rm for}~
 M_\sigma > \min \left[\Gamma_\chi, \Gamma_\phi \right]\,.
\end{eqnarray}

\bigskip\bigskip\bigskip
In addition to the above requirements, we require that the reheating temperature of the curvaton $T_{\rm reh}^{(\sigma)}$ be higher than the spharelon temperature $T_{\rm spharelon} = {\cal O}(100)~{\rm GeV}$ in order that leptogenesis works. We show contours of $T_{\rm reh}^{(\sigma)}$ in Figs.~\ref{fig:const_Imw_r} and \ref{fig:const_Imw_Msigma}. Actually, in the parameter space shown in these figures, the entire regions satisfy $T_{\rm reh}^{(\sigma)}> T_{\rm spharelon}$ and hence no region is excluded from this requirement.

Note that, in general, baryon isocurvature fluctuations can be generated in our scenario since baryon asymmetry is created from the decay of the curvaton.
When the curvaton gives a dominant contribution to the total curvature perturbation, the amplitude of baryon isocurvature fluctuations is proportional to $(1 - r_{\rm dec}) /r_{\rm dec}$ (see, e.g., Ref.~\cite{Lyth:2002my}). However, in the mixed case, the curvature perturbation from the inflaton dilutes the isocurvature fluctuations. Moreover, from the constraint on the primordial non-Gaussianity, we consider the case with $r_{\rm dec} \simeq 1$,
and hence the baryon isocurvature fluctuations should be completely negligible in our scenario, which is consistent with the current observational constraint.

\bigskip\bigskip\bigskip
Now we discuss the implications of the results shown in Figs.~\ref{fig:const_Imw_r} and \ref{fig:const_Imw_Msigma}. From Fig.~\ref{fig:const_Imw_r} where the constraints are shown in the $\Im \omega$--$r$ plane for several fixed values of $M_\sigma$, one can see that when the mass of $\sigma$ is large as $M_\sigma =10^{10}~{\rm GeV}$ or small as $M_\sigma = 10^5~{\rm GeV}$, allowed region becomes very small, i.e.,  we do not have a parameter range which satisfies all the constraints/requirements. This can be understood as follows. 
When $M_\sigma$ is large as $M_\sigma=10^{10}~{\rm GeV}$ (left panel of Fig.~\ref{fig:const_Imw_r}), $\Gamma_\sigma / M_\sigma$ (for $M_\sigma < \min [\Gamma_\chi, \Gamma_\phi]$) or
$\Gamma_\sigma$ (for $M_\sigma > \min [\Gamma_\chi, \Gamma_\phi]$) gets high. In either case, the  initial curvaton amplitude $\sigma_\ast$ should be large in order to have $\rdec \sim 1$ which is demanded from the non-Gaussinaity constraint. However, as seen from Eq.~\eqref{eq:R}, larger $\sigma_\ast$ indicates smaller $R$, which drives the tensor-to-scalar ratio bigger. Therefore the excluded region is pushed up towards a higher value of $r$, which makes the allowed region smaller. On the other hand, when $M_\sigma$ is small as $M_\sigma=10^5~{\rm GeV}$ (right panel of Fig.~\ref{fig:const_Imw_r}), the reheating temperature $T_\sigma$ gets smaller, which also decreases the baryon asymmetry $n_B/s$, and hence the constraint on $\Im \omega$ becomes severer and the excluded region extends to a higher value of $\Im \omega$. Therefore the regions satisfying all the constraints tend to diminish. 

In Fig.~\ref{fig:const_Imw_Msigma}, the allowed region in the $\Im \omega$--$M_\sigma$ plane is shown for some fixed values of $r$, from which one can see that as $r$ decreases, the regions with all the constraints satisfied become smaller and eventually disappear.  This can be intuitively understood as follows.
First, from Eq.~\eqref{eq:baryon_reheat}, one can find that, to account for the present value of the baryon asymmetry, a relatively higher reheating temperature of the curvaton, that is,
a larger decay rate of the curvaton is needed.
On the other hand, to be consistent with the current constraint on the primordial non-Gaussianity, the curvaton should almost dominate the energy density of the Universe before its decay, which requires a smaller decay rate for the curvaton. Since the decay rate of the curvaton is a function of $M_\sigma$ and $\Im \omega$, the allowed region in the parameter space is constrained as shown in Fig.~\ref{fig:const_Imw_Msigma}.
For the case with a smaller tensor-to-scalar ratio, which corresponds to a lower energy scale of inflation,
the required upper bound on the decay rate of the curvaton becomes more stringent.
Eventually, for $r \lesssim 10^{-4}$ it is difficult to realize a high enough reheating temperature for the baryon asymmetry.

The results presented in Figs.~\ref{fig:const_Imw_r} and \ref{fig:const_Imw_Msigma} suggest that, in order that sneutrino inflation and leptogenesis scenario work simultaneously, the mass of the lightest sneutrino should be in the range of $10^5~{\rm GeV} \lesssim M_\sigma \lesssim 10^{10}~{\rm GeV}$ and the tensor-to-scalar ratio $r$ cannot be smaller than $10^{-4}$.
As shown in Fig.~\ref{fig:const_Imw_Msigma},  when $r=0.01$, the allowed region in the $\Im\omega$--$M_\sigma$ plane is relatively large, and the lightest sneutrino mass can be $10^5~{\rm GeV} \lesssim M_\sigma \lesssim 10^{10}~{\rm GeV}$ whose actual range depends on $\Im\omega$. However, when $r=10^{-4}$, the allowed range disappears, which means that, if this scenario describes the inflationary dynamics and is responsible for the baryon asymmetry of the Universe, the tensor-to-scalar ratio should be $r> 10^{-4}$, which could be accessible in future CMB B-mode observations such as LiteBIRD \cite{Hazumi:2021yqq} or in an ideal case of 21cm fluctuations \cite{Book:2011dz}.

Finally, we comment on how the assumption on the lightest neutrino mass affects our results. In the analysis for Figs.~\ref{fig:const_Imw_r} and \ref{fig:const_Imw_Msigma}, we have fixed the lightest neutrino mass as $m_{\rm lightest} = 10^{-5}~{\rm eV}$.
From the expressions \eqref{eq:decay_sigma} - \eqref{eq:epsilon_1}, 
for the case with $\Im \omega \gtrsim 1$,
the decay rate of the sneutrinos and the CP-violation parameter are almost independent of $m_{\rm lightest}$ as long as 
$m_1 \ll m_2$.
On the other hand, for the case with the relatively small $\Im \omega$,
the CP-violation parameter and also the decay rate of the sneutrinos 
become dependent on $m_{\rm lightest}$. To see how $m_{\rm lightest}$ affects our results, in Appendix~B, we show the allowed regions in the
$\Im \omega$--$r$ plane for some different values of $m_{\rm lightest}$. We also show the result in the $m_{\rm lightest}$--$M_\sigma$ plane. As shown in the figures in Appendix~B, 
the main conclusion that $10^5~{\rm GeV} \lesssim M_\sigma \lesssim 10^{10}~{\rm GeV}$ and $r \gtrsim 10^{-4}$ should be satisfied still holds without much regard to the choice of $m_{\rm lightest}$. 

In this section, we have focused on the case with normal mass hierarchy. As we have already mentioned, even if we consider the inverted mass hierarchy, our results do not change qualitatively. However, the actual values allowed from the requirements (i)-(iv) slightly change, which we show in Appendix~C.

\section{Conclusions and Discussion} \label{sec:conclusion}

We have argued that the sneutrino inflation scenario works without conflicting with the current observational constraints on $n_s$ and $r$ when two heavier of them play a role of the inflatons and the lightest one as the curvaton. We have also shown that a successful baryogenesis scenario is realized in the same framework through leptogenesis via the lightest sneutrino decay.

Since sneutrinos have a quadratic potential, such a model does not work if only a single field is responsible for the inflationary dynamics and primordial fluctuations. However, it has been shown that if three scalar fields with all having a quadratic potential play a role of two-inflaton and one-curvaton, such a model can be consistent with the current observational bounds on $n_s$ and $r$ \cite{Morishita:2022bkr}. We have applied this scenario to the sneutrino framework and investigated the parameter range consistent with observations. We have simultaneously investigated whether a successful baryogenesis scenario via sneutrino leptogenesis can work in this setup. In Figs.~\ref{fig:const_Imw_r} and \ref{fig:const_Imw_Msigma}, we have presented the parameter space which is consistent with observations of primordial fluctuations and can realize a right amount of baryon asymmetry of the Universe. We have found that, for a successful scenario, the lightest sneutrino mass should be in the range of $10^5~{\rm GeV} \lesssim M_\sigma \lesssim 10^{10}~{\rm GeV}$ and the tensor-to-scalar ratio should be $r \gtrsim 10^{-4}$, the latter of which could be detectable in future observations.  

As a final remark, we comment on dark matter candidates in the model. Since we do not specify the supersymmetry breaking scenario, we cannot discuss dark matter candidates and their relic abundance in a concrete manner. Nevertheless, the curvaton reheating temperature $T_{\rm reh}^{(\sigma)}$ is calculable in the model, and the value of $T_{\rm reh}^{(\sigma)}$ can provide a constraint on feebly-interacting dark matter candidates.
However, finding a dark matter candidate in our framework would be relatively easy since there is a  wide range of possibilities in supersymmetric models. Given that the sneutrino scenario presented in this paper is attractive in a sense that we can describe the inflationary universe and primordial fluctuations as well as the baryon asymmetry of the Universe simultaneously, and hence extending a model to include a dark matter candidate would be an interesting avenue, which we will come back in the future.

\section*{Acknowledgements}
We would like to thank Masahide Yamaguchi for useful comments on baryon isocurvature fluctuations.
The authors thank the Yukawa Institute for Theoretical Physics at Kyoto University, where this work was initiated during the YITP-T-21-08 on ``Upcoming CMB observations and Cosmology".
This work is supported in part by JSPS KAKENHI Grant Number 19K03874 (TT), 19K147101 (TY), JP20H01932 (SY) and JP20K03968
(SY).  

\appendix
\pagebreak
\def\thesection{Appendix: \Alph{section}}
\section{\label{sec:appA} K\"ahler potential and superpotential}
\def\thesection{\Alph{section}}

We consider a K\"ahler potential in no-scale supergravity~\cite{Cremmer:1983bf,Ellis:1984bf} so that supergravity effects do not hinder chaotic inflation.
Here the superpotential takes the same form as in the global supersymmetry case, which allows us to interconnect the inflation and the seesaw model.
The K\"ahler potential in our model is
\begin{align} 
K &= -3 \, \log\left[ T+T^\dagger
-\frac{1}{3}\left(\hat{n}_\phi^\dagger\hat{n}_\phi+\hat{n}_\chi^\dagger\hat{n}_\chi+\hat{n}_\sigma^\dagger \hat{n}_\sigma\right)\right]+\Phi_{MSSM}^\dagger \,e^V \Phi_{MSSM} \,.
\label{kaehler}
\end{align}
In this appendix, we adopt the unit where the reduced Planck mass $M_{\rm pl}=1$.
Here $T$ is a modulus superfield, $\hat{n}_\phi, \hat{n}_\chi, \hat{n}_\sigma$ are the singlet neutrino chiral superfields,
and $\Phi_{MSSM}$ and $V$ collectively denote the chiral superfields and the gauge superfields of the minimal supersymmetric Standard Model (MSSM), respectively. 
The superpotential is given by
\begin{align} 
W &= W_{MSSM}+\frac{1}{2}\hat{M}_\phi\,\hat{n}_\phi^2+\frac{1}{2}\hat{M}_\chi\, \hat{n}_\chi^2+\frac{1}{2}\hat{M}_\sigma\,\hat{n}_\sigma^2
+\hat{h}_{\phi\alpha} \, \hat{n}_\phi H_u L_\alpha
+\hat{h}_{\chi\alpha} \, \hat{n}_\chi H_u L_\alpha
+\hat{h}_{\sigma\alpha} \, \hat{n}_\sigma H_u L_\alpha,
\label{sup}
\end{align}
where $W_{MSSM}$ is the MSSM superpotential, $\hat{M}_\phi, \hat{M}_\chi, \hat{M}_\sigma$ are supersymmetric masses for the singlet neutrinos, and $\hat{h}_{\phi\alpha},\hat{h}_{\chi\alpha},\hat{h}_{\sigma\alpha}$ are the neutrino Dirac Yukawa couplings with $\alpha=e,\mu,\tau$\footnote{
One can further introduce terms that induce supersymmetry breaking and set the present cosmological constant.
}.
Note that the K\"ahler potential Eq.~(\ref{kaehler}) has the following shift symmetry:
\begin{align}
    T \to T+\frac{1}{3}\left(\epsilon_\phi^* \hat{n}_\phi +\epsilon_\chi^* \hat{n}_\chi +\epsilon_\sigma^* \hat{n}_\sigma\right), \ \ \hat{n}_\phi\to \hat{n}_\phi+\epsilon_\phi, \ \ \hat{n}_\chi\to \hat{n}_\chi+\epsilon_\chi, \ \ \hat{n}_\sigma\to \hat{n}_\sigma+\epsilon_\sigma \,.
\end{align}
Here $\epsilon_\phi,\epsilon_\chi,\epsilon_\sigma$ are infinitesimal parameters.  This symmetry is explicitly broken by the sneutrino superpotential in Eq.~(\ref{sup}).

The action responsible for the inflation and the seesaw mechanism is given by
\begin{align} 
S &\supset \int{\rm d}^4x \sqrt{-g}
\left[ \ K_{i\bar{j}} \, \partial_\mu \hat{n}_j^\dagger \, \partial^\mu \hat{n}_i + K_{i\bar{j}} \, \bar{\psi}_{\hat{n}_j} i\bar{\sigma}^\mu \partial_\mu\psi_{\hat{n}_i} 
+K_{i\bar{T}} \, \partial_\mu T^\dagger \partial^\mu \hat{n}_i + K_{i\bar{T}} \, \bar{\psi}_{T} i\bar{\sigma}^\mu \partial_\mu\psi_{\hat{n}_i}+{\rm H.c.}\right.
\nonumber \\
&+K_{T\bar{T}} \, \partial_\mu T^\dagger \partial^\mu T + K_{T\bar{T}} \, \bar{\psi}_{T} i\bar{\sigma}^\mu \partial_\mu\psi_{T}
\nonumber \\
&-e^{K/2}\left\{ \, \frac{1}{2}\hat{M}_i \, \psi_{\hat{n}_i}^Ti\sigma_2\psi_{\hat{n}_i} 
+\hat{h}_{i\alpha} \, \psi_{\hat{n}_i}^Ti\sigma_2 H_u \psi_{L_\alpha}+\hat{h}_{i\alpha} \, \hat{n}_i\, \psi_{H_u}^Ti\sigma_2 \psi_{L_\alpha}
+\hat{h}_{i\alpha} \, \psi_{\hat{n}_i}^Ti\sigma_2 \psi_{H_u} L_\alpha
+{\rm H.c.} \, \right\}
\nonumber \\
&-e^K \left\{ \, (K^{-1})_{i\bar{j}}(W_i+K_iW)(W_j+K_jW)^\dagger+(K^{-1})_{i\bar{T}}(W_i+K_iW)(K_TW)^\dagger+{\rm H.c.}\right.
\nonumber \\
&\left.\left. +(K^{-1})_{T\bar{T}}(K_TW)(K_TW)^\dagger-3\vert W\vert^2 \, \right\} \ \right],
\label{actionpre}
\end{align}
where $\psi$'s represent the fermionic components, and we have defined
\begin{align}
&K_i\equiv\frac{\partial K}{\partial \hat{n}_i},\ \ K_T\equiv\frac{\partial K}{\partial T},
 \ \ K_{i\bar{j}}\equiv \frac{\partial^2 K}{\partial \hat{n}_i \partial \hat{n}_j^\dagger},
\ \ K_{i\bar{T}}\equiv\frac{\partial^2 K}{\partial \hat{n}_i \partial T^\dagger}, \ \
K_{T\bar{T}}\equiv \frac{\partial^2 K}{\partial T \partial T^\dagger}, \ \ 
W_i\equiv\frac{\partial W}{\partial \hat{n}_i},
\nonumber
\end{align}
 and $i,j$ are indices that run as $i,j=\phi,\chi,\sigma$. Contractions over $i,j$ are understood.
The derivatives of the K\"ahler potential are calculated as
\begin{align} 
K_{i\bar{j}}&=e^{2K/3}\left[\delta_{ij}\left\{T+T^\dagger - \frac{1}{3}\left(\hat{n}_\phi^\dagger\hat{n}_\phi+\hat{n}_\chi^\dagger\hat{n}_\chi+\hat{n}_\sigma^\dagger\hat{n}_\sigma\right)\right\}+\frac{1}{3}\hat{n}_i^\dagger\hat{n}_j\right],
\nonumber\\
K_{i\bar{T}}&=-e^{2K/3}\hat{n}_i^\dagger,
\ \ \
K_{T\bar{T}}=3e^{2K/3},
\nonumber
\end{align}
 and the scalar potential reduces to the following simple form thanks to the no-scale structure:
\begin{align} 
&e^K \left\{ \, (K^{-1})_{i\bar{j}}(W_i+K_iW)(W_j+K_jW)^\dagger+(K^{-1})_{i\bar{T}}(W_i+K_iW)(K_TW)^\dagger+{\rm H.c.}\right.
\nonumber \\
&\left. +(K^{-1})_{T\bar{T}}(K_TW)(K_TW)^\dagger-3\vert W\vert^2 \, \right\}
\nonumber \\
&=e^{2K/3}\left(\left\vert\frac{\partial W}{\partial \hat{n}_\phi}\right\vert^2+\left\vert\frac{\partial W}{\partial \hat{n}_\chi}\right\vert^2+\left\vert\frac{\partial W}{\partial \hat{n}_\sigma}\right\vert^2\right).
\end{align}

Now the K\"ahler potential Eq.~(\ref{kaehler}) is assumed to develop a VEV as
\begin{align} 
\langle K \rangle &=c .
\end{align}
Also, the VEVs of the sneutrinos are assumed to satisfy the following relations at all times:
\begin{align} 
e^{-c/3} &\gg \vert\langle\hat{n}_i\rangle\vert,
\ \ \ \ \ 
e^{-c/3} \gg \frac{1}{3}\vert\langle\hat{n}_i\rangle\vert\vert\langle\hat{n}_j\rangle\vert.
\ \ \ \ \ \  \ \ (i,j=\phi,\chi,\sigma)\,(i\neq j)
\label{vevhierarchy}
\end{align}
Given Eq.~(\ref{vevhierarchy}), we can neglect kinetic mixings among the sneutrinos and the modulus, and make an approximation with $K_{i\bar{j}}\simeq e^{c/3}\,\delta_{ij}$ and define canonically-normalized singlet neutrino superfields as $\phi=e^{c/6}\,\hat{n}_\phi$, $\chi=e^{c/6}\,\hat{n}_\chi$, $\sigma=e^{c/6}\,\hat{n}_\sigma$.
In terms of the canonically-normalized fields, and with the redefinition of masses and Yukawa couplings $M_i=e^{c/6}\,\hat{M}_i$, $h_{i\alpha}=e^{c/3}\,\hat{h}_{i\alpha}$ ($i=\phi,\chi,\sigma$),
the part of the action Eq.~(\ref{actionpre}) involving the neutrinos and sneutrinos is recast into Eq.~(\ref{action})\footnote{
The lepton $\psi_{L_\alpha}$ and the slepton $L_\alpha$ in Eq.~(\ref{actionpre}) are rewritten as $L_\alpha$ and $\widetilde{L}_\alpha$ in Eq.~(\ref{action}), respectively.
}.

The K\"ahler potential Eq.~(\ref{kaehler}) receives quantum corrections from the sneutrino superpotential in Eq.~(\ref{sup}) that breaks the no-scale structure, and these quantum corrections might spoil the chaotic inflation scenario.
However, one can check that such quantum corrections are sufficiently small that chaotic inflation still works, as has been done in Section~4 of Ref.~\cite{Haba:2017fbi}. We leave this check for future work.

\pagebreak
\def\thesection{Appendix: \Alph{section}}
\section{\label{sec:appB} 
$m_{\rm lightest}$ dependence}
\def\thesection{\Alph{section}}

\begin{figure}[H]
\centering
  \includegraphics[width=16cm]{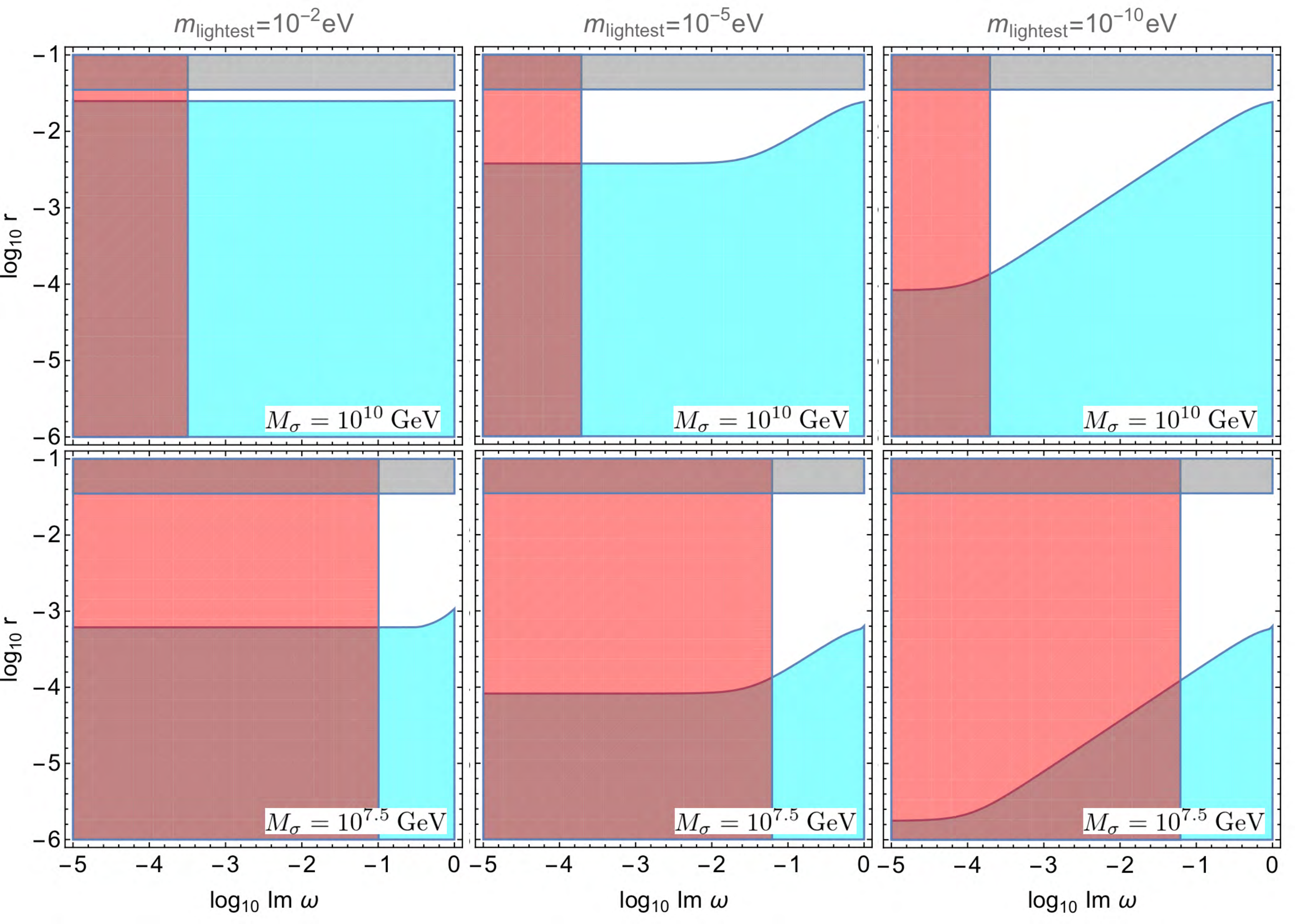}
\caption{\label{fig:const_Imw_r_s_2} 
Constraints on the $\Im \omega$ - $r$ plane for the cases of $m_{\rm lightest} = 10^{-2}~{\rm eV}$ (left panels),  $10^{-5}~{\rm eV}$ (middle panels) and $10^{-10}~{\rm eV}$ (right panels). The mass of the curvaton is taken as $M_\sigma = 10^{10}~{\rm GeV}$ (top row) and $10^{7.5}$~${\rm GeV}$ (bottom row). 
The meaning of shading by colors is the same as in Fig.~\ref{fig:const_Imw_r}. 
2018 result. The magenta region is excluded by the constraint from the perturbativity on the Yukawa coupling. The cyan region is excluded by the requirement that $\rdec = 1$ coming from the constraint on the primordial non-Gaussianity. The red region is that inaccessible to account for the present value of the baryon-to-photon ratio. 
In all panels, the whole region satisfy $T_{\rm reh}^{(\sigma)} > 10^4~{\rm GeV}$.}
\end{figure}

In this appendix, we show more plots for the normal mass hierarchy case. In the main text, we have fixed the lightest neutrino mass as $m_{\rm lightest} = 10^{-5}~{\rm eV}$. In Fig.~\ref{fig:const_Imw_r_s_2}, we show the same plot as Fig.~\ref{fig:const_Imw_r}, but for the cases with $m_{\rm lightest} = 10^{-2}~{\rm eV}$ and $10^{-10}~{\rm eV}$ along with the case of $m_{\rm lightest}=10^{-5}~{\rm eV}$.  In the figure, the value of $\Im \omega$ is shown in the logarithmic scale, and hence we can see the behavior in the small $\Im\omega$ region more clearly. 

\begin{figure}[H]
\centering
  \includegraphics[width=16cm]{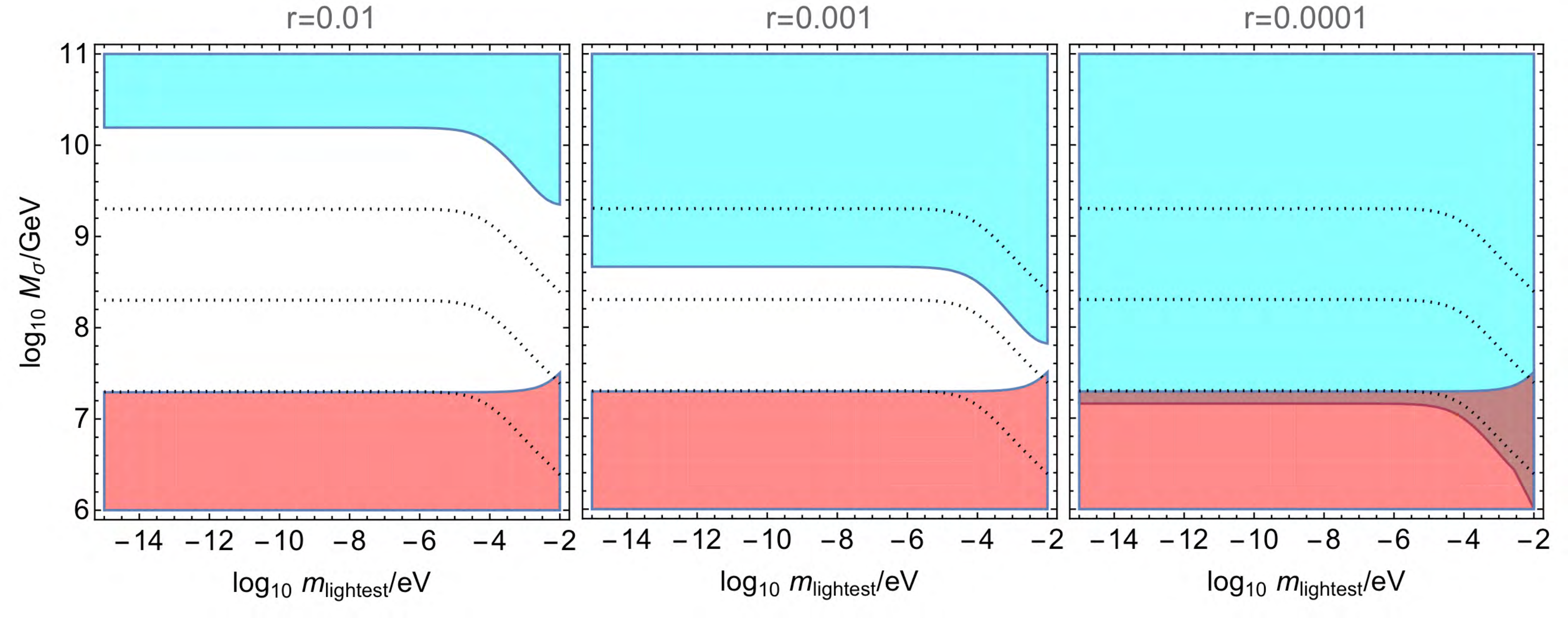}
\caption{\label{fig:const_lightest_Msigma} Constraints on $m_{\rm lightest}$ - $M_\sigma$ plane for $\Im \omega =0.1$. From the left to the right, we change the tensor-to-scalar ratio as $r=0.01$, $0.001$, 
and $0.0001$. 
The meaning of shading by colors is the same as in Fig.~\ref{fig:const_Imw_r}. 
that $\rdec = 1$ coming from the constraint on the primordial non-Gaussianity. 
In each panel, the black-dotted lines are those of the reheating temperature of the curvaton, $T_{\rm reh}^{(\sigma)} = 10^9$, $10^8$, and $10^7$~GeV from top to bottom.}
\end{figure}

In Fig.~\ref{fig:const_lightest_Msigma}, we show the constraints in the $m_{\rm lightest}$--$M_\sigma$ plane. In the figure, we fix $\Im\omega$ as $\Im\omega=0.1$ and take several values for $r$ as shown above each panel.
From Figs.~\ref{fig:const_Imw_r_s_2} and \ref{fig:const_lightest_Msigma}, the overall behaviour of the allowed regions are almost the same when $m_{\rm lightest} \ll {\cal O}(0.01)~{\rm eV}$, although the actual range of allowed values for the model parameters can change for relatively large $m_{\rm lightest}$.

\pagebreak
\def\thesection{Appendix: \Alph{section}}
\section{\label{sec:appC} Case with inverted mass hierarchy}
\def\thesection{\Alph{section}}

\begin{figure}[H]
\centering
  \includegraphics[width=16cm]{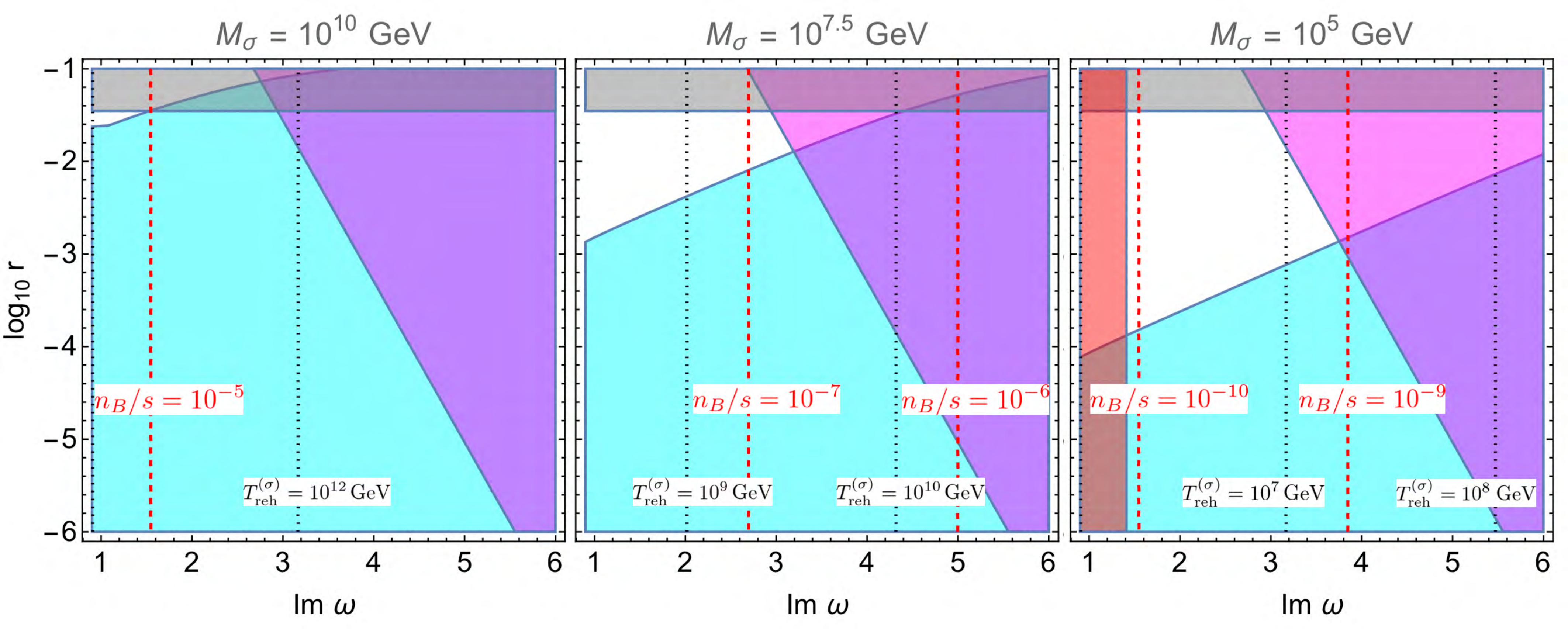}
\caption{\label{fig:const_Imw_r_inv} Constraints on $\Im \omega$ - $r$ plane for inverted hierarchy. From the left to the right, we change the mass of the curvaton as $M_\sigma = 10^{10}$, $10^{7.5}$, and $10^5$~${\rm GeV}$. 
The meaning of shading by colors is the same as in Fig.~\ref{fig:const_Imw_r}. 
The vertical red dashed lines are corresponding to the contours of the baryon-to-entropy ratio, $n_B / s$, and the black-dotted lines represent the reheating temperature of the curvaton, $T_{\rm reh}^{(\sigma)}$. 
}
\end{figure}

\begin{figure}[H]
\centering
  \includegraphics[width=16cm]{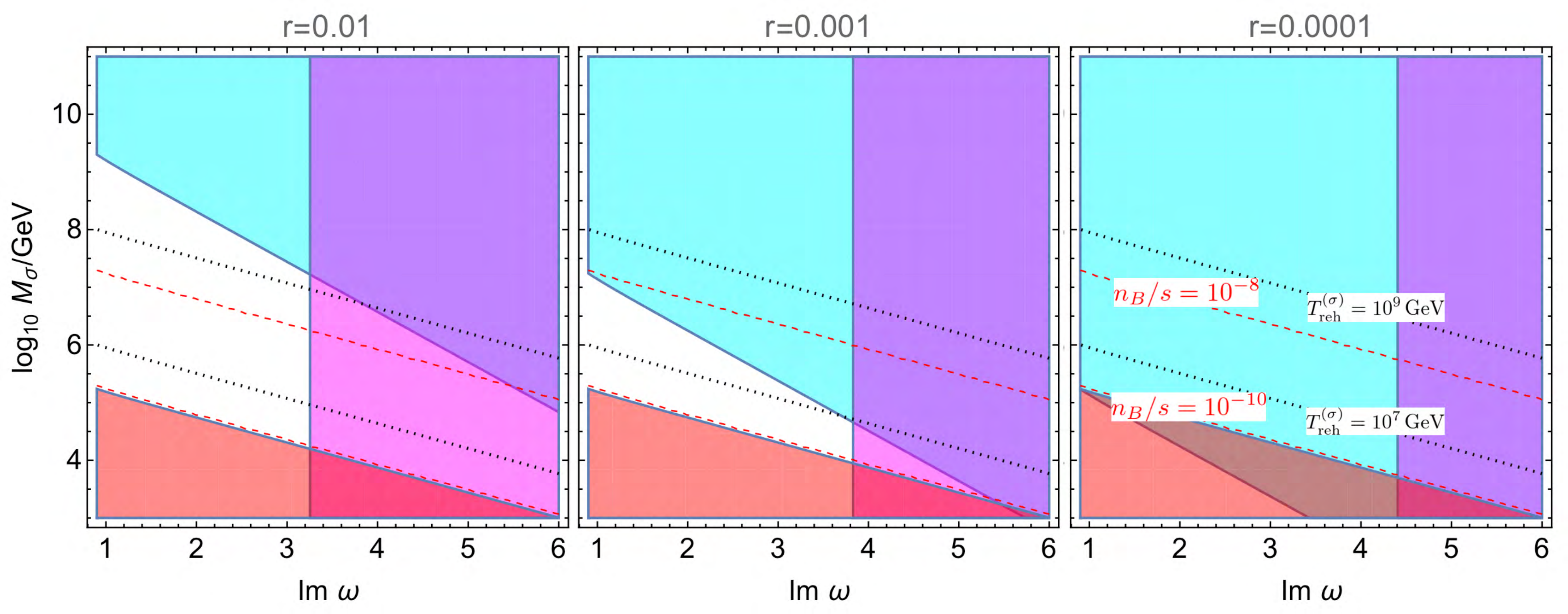}
\caption{\label{fig:const_Imw_Msigma_invert} Constraints on the $\Im \omega$ - $M_\sigma$ plane for inverted hierarchy. From left to right, we change the tensor-to-scalar ratio as $r=0.01$,$0.001$, and $0.0001$. 
The meaning of shading by colors is the same as in Fig.~\ref{fig:const_Imw_r}. 
In each panel, the red dashed (black dotted) lines correspond to $n_B/s = 10^{-8},$ and $10^{-10}$, ($T_{\rm reh}^{(\sigma)} = 10^9~{\rm GeV},$ and $10^{7}~{\rm GeV}$) from the top to the bottom.}
\end{figure}

\begin{figure}[H]
\centering
  \includegraphics[width=16cm]{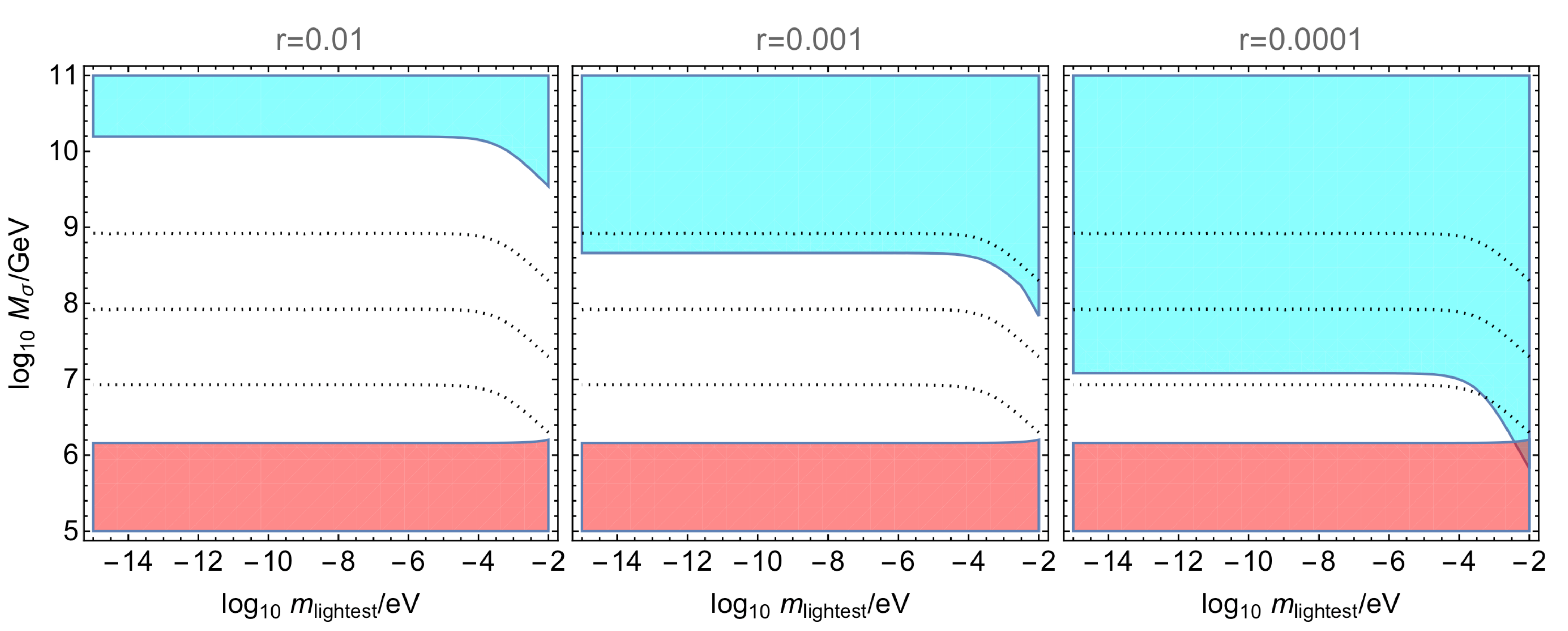}
\caption{\label{fig:const_lightest_Msigma_inv} Constraints on the $m_{\rm lightest}$ - $M_\sigma$ plane for inverted hierarchy. From left to right, we change the tensor-to-scalar ratio as $r=0.01$, $0.001$, and $0.0001$. 
The meaning of shading by colors is the same as in Fig.~\ref{fig:const_Imw_r}. 
In each panel, the black-dotted lines represent  the reheating temperature of the curvaton, $T_{\rm reh}^{(\sigma)} = 10^9$, $10^8$, and $10^7$~GeV.}
\end{figure}

Here we present our results for 
the inverted mass hierarchy case. In Fig.~\ref{fig:const_Imw_r_inv} (and \ref{fig:const_Imw_Msigma_invert}), we show excluded regions from (i)-(iv) in Sec.~\ref{sec:results} in the $\Im\omega$--$r$ plane (and $\Im\omega$--$M_\sigma$ plane) which can be compared to the one for the normal mass hierarchy case shown in Fig.~\ref{fig:const_Imw_r} (and \ref{fig:const_Imw_Msigma}). As one can see from the figure, a general tendency is the same as that in the normal mass hierarchy case, however, the range of $M_\sigma$ allowed by the constraints becomes larger in the inverted mass hierarchy case. In Fig.~\ref{fig:const_lightest_Msigma_inv}, the constraints are shown in the $m_{\rm lightest}$--$M_\sigma$ plane which can be compared to Fig.~\ref{fig:const_lightest_Msigma} for the normal hierarchy case. Again one can see that the inverted mass hierarchy case allows more parameter range for the successful scenario. 

\pagebreak
\bibliography{refs}

\end{document}